 \documentclass[aps,preprint,superscriptaddress]{revtex4-1}
  \usepackage[T1]{fontenc}
  \usepackage[english]{babel}
  \usepackage{times}
  \usepackage{amssymb,amsmath,amsfonts,amsthm}
  \usepackage{mathtools}
  \usepackage{verbatim}
  \usepackage{enumerate}
  \usepackage{graphicx}
  \usepackage{hyperref}
  \usepackage{bbm}
  \usepackage{booktabs}
  
  \usepackage[caption=false]{subfig}
  \usepackage{mathrsfs}
  \usepackage{color}
  \usepackage{bm}
  \usepackage{braket}
  \usepackage{euscript}
\usepackage{graphicx}
\usepackage{multirow}
\usepackage{bm}

\usepackage{etoolbox}
\makeatletter
\patchcmd{\frontmatter@abstract@produce}
  {\vskip200\p@\@plus1fil
   \penalty-200\relax
   \vskip-200\p@\@plus-1fil}
  {}
  {}
  {} 
\makeatother
\begin{document}

\title{Quantum simulation beyond Hamiltonian paradigm:
categorical quantum simulation}
\author{Yuanye Zhu}
\thanks{Corresponding author: zhuyy16@mails.tsinghua.edu.cn}
\affiliation{State Key Laboratory of Low-dimensional Quantum Physics, Beijing, 100084, China}
\affiliation{Department of Physics, Tsinghua University, Beijing, 100084, China}

\begin{abstract}
With the development of topological field theory, the mathematical tool of the tensor category was also introduced into physics. Traditional group theory corresponds to a special category,group category. Tensor categories can describe higher-order interactions and symmetric relations, while group theory can only describe first-order interactions. In fact, the quantum circuit itself constitutes a category. However, at present, the field of quantum computing mainly uses group theory as a mathematical tool. If category theory is introduced into the field of quantum simulation, the application scope of quantum computers can be greatly expanded. This paper propose  a new dynamic simulation method,categorical quantum simulation. In our paradigm quantum simulation is no longer based on the structure of the  group theory, but based on the structure of the tensor category. This could enable many systems that could not be efficiently quantum simulated before.In this article we give an concrete example of the categorical simulation of $SU(3)$ Yang-Mills theory. It shows that categorical quantum simulation provides a new encoding method,emergenism encoding, which saves more qubits resources than reductionism quantum encoding. In addition, many domains can be described in the language of category theory, which allows quantum circuits to directly encode and simulate these domains. 
\end{abstract}

\maketitle

\paragraph{Introduction}
Even today, the question of whether quantum systems can simulate effectively remains challenging. Because as the number of degrees of freedom, or particles, increases, the computer savings required for storage systems are enormous, and these storage requirements increase exponentially with degrees of freedom\cite{Feynman1982SimulatingPW}. Moreover, in order to simulate a quantum system, it is necessary to simulate its time evolution, and the operators needed to simulate time evolution increase exponentially with the size of the system. This exponential increase is inevitable unless some approximation method like Monte Carlo is used\cite{Kroese2014WhyTM,Metropolis1949TheMC,Rubinstein1981SimulationAT}. 

One proposal was given by  Richard Feynman in 1982 as Hamiltonian simulation\cite{Feynman1982SimulatingPW}. He pointed out that let the computer itself be built of quantum mechanical elements which obey quantum mechanical laws. Feynman realized that quantum computers themselves had undergone an exponential explosion in evolution. Because quantum computers can process exponentially large amounts of physical information without having to use the storage resources of exponential explosions.  This makes it a natural tool for efficient quantum simulations. Feynman points out that one can use controlled quantum systems to simulate more complex quantum systems. One just need to find the map between the initial state, final state, and Hamiltonian evolution of these two systems. Specifically, for a simulated quantum system $|\phi\rangle$, which  evolves from the initial state  $|\phi\left(0\right)\rangle$ to $|\phi\left(t\right)\rangle$ through the operator $U=\exp \left\{-i \hbar H_{\text {sys }} t\right\}$. Where $ H_{\text {sys }}$ is the Hamiltonian of simulated system. For a quantum simulator $|\psi\rangle$, it is a controllable quantum system whose initial state is $|\psi\left(0\right)\rangle$ that can be prepared. Its evolution operator is $U'=\exp \left\{-i \hbar H_{\text {sim }} t\right\}$, in which $ H_{\text {sim }}$ is  the Hamiltonian of quantum simulator that can be controlled (or programmable). The final state is $|\psi\left(t\right)\rangle$ that is measurable. If the mapping between the simulator and system (between  $|\psi\left(0\right)\rangle$ and  $|\phi\left(0\right)\rangle$, between  $|\psi\left(t\right)\rangle$ and  $|\phi\left(t\right)\rangle$) exit, then the system can be simulated\cite{Abrams1997SimulationOM,Lidar1997SIMULATINGIS,Lloyd1996UniversalQS,
Marzuoli2002SpinNQ,Ortiz2001QuantumAF,Raeisi2012QuantumcircuitDF,Marzuoli2002SpinNQ,Terhal2000ProblemOE,
Verstraete2008QuantumCF,Wiesner1996SimulationsOM,Zalka1996EfficientSO,
Zalka1998SimulatingQS}. 
 \begin{figure}
  \begin{center}
 \includegraphics[width=0.5\textwidth]{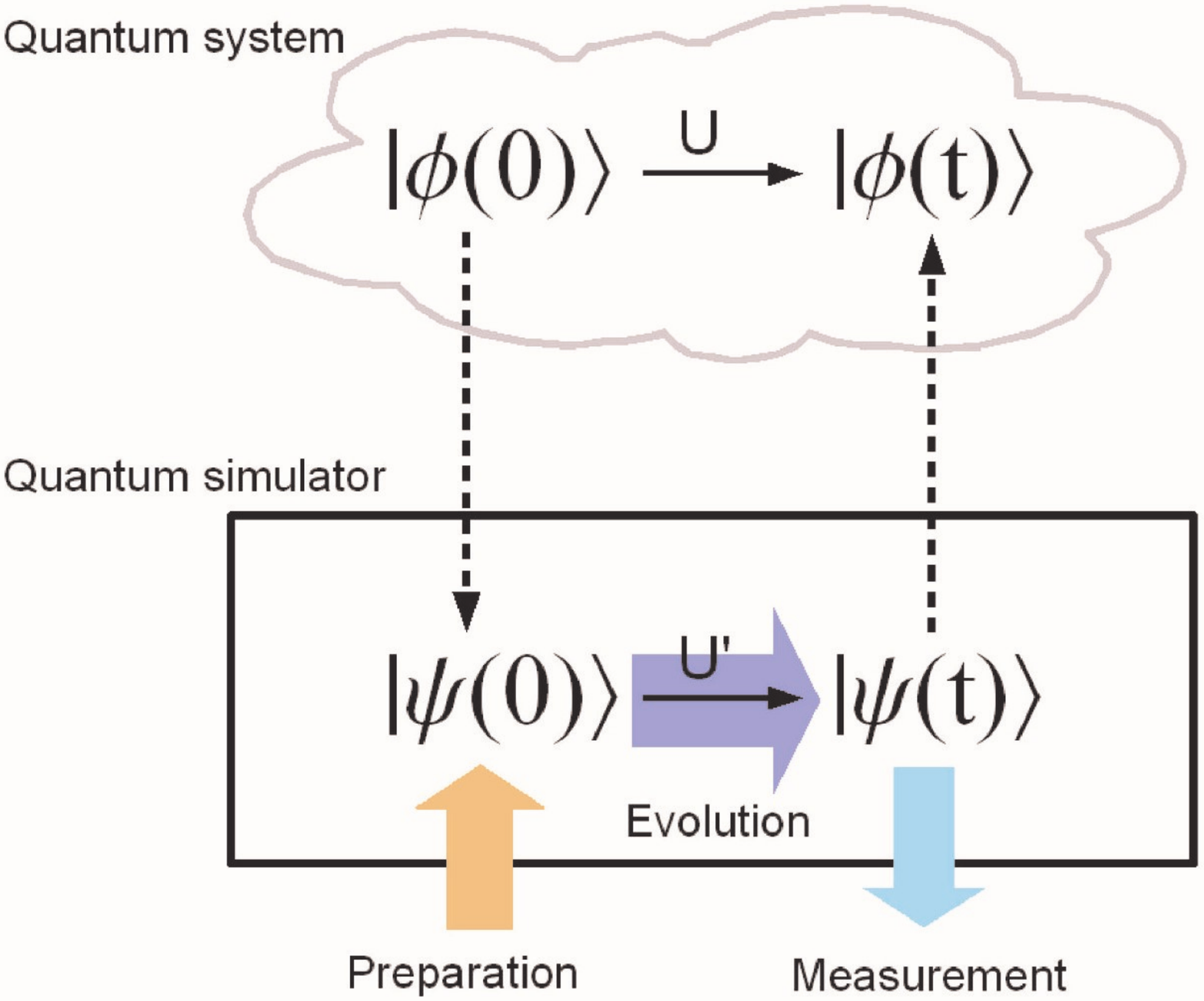}
  \caption{Feynman's quantum simulation paradigm\cite{RevModPhys.86.153}   (or Hamiltonian simulation) uses a controllable quantum simulator $|\phi\rangle$ to simulate a complex or un-controllable quantum system $|\psi\rangle$. }\label{fig1}
  \end{center}
  \end{figure}

But think carefully about Feynman's paradigm, Hamiltonian simulation, which does not maximize the use of resources such as quantum computer parallelism in quantum simulations. Because it has some scope. The Feynman paradigm requires that the simulated system must be a quantum system, which greatly narrows the range of quantum simulations that quantum computers can perform. In fact, many classical systems such as logic, topology, and even optimization problems require powerful arithmetic to solve. These problems cannot be described by convincing quantum states or Hamiltonian quantities. Effective quantum simulation of these problems requires the search for new paradigms of quantum simulation that go beyond Feynman's paradigm. The reason for this is that the Feynman paradigm can simulate a system that has the structure of group. Specifically, the evolution operator of the simulated system forms the structure of a $SU(2^n)$ lie group, and the quantum simulator is also the structure of a $SU(2^N)$ lie group, so it is only necessary to find the group homomorphic mapping to complete the quantum simulation. In fact, this is one of the limitations of the Feynman paradigm. Group homomorphism mapping is not easy to come by because not all systems, or even most systems, evolve like simple group structures. So breaking the Feynman paradigm and expanding the range of quantum computer simulations requires the introduction of new mathematical theories.

In fact, most Feynman quantum simulations are space discrete, using lattice models to encode the states of the simulated systems into quantum states, which, while intuitive, is a waste of qubit sources. In fact, there are many times when we don't need complete state information, so we don't need to use such a large computational resource for quantum computing. This is one of the shortcomings of Feynman's paradigm. Actually, when we know a little more about the mathematical structure behind the simulated system, we can find more concise coding that will save qubit resources. This is like saying that in quantum information there is very little need to solve schrodinger's equation directly, instead we just need to know the Hamiltonian $H$ and initial state $|\psi\rangle$, and then we write it as $e^{-i  \hbar H t}|\psi\rangle$, which must be evolved quantum states at time $t$. This direct algebraic multiplication greatly reduces operational resources compared to solving differential equations. Why are we allowed to do this? It is because we have mastered the algebraic structure behind quantum mechanics. Since Schrodinger's equation, its temporal evolution operator, forms the structure of lie group, it must be legal and correct for people to write the quantum state at time $t$ directly. So what we need to do is to find this algebraic structure for the simulated system and quantum simulator, so that it can be encoded efficiently and better use quantum computing resources.

\begin{figure}
  \begin{center}
 \includegraphics[width=0.4\textwidth]{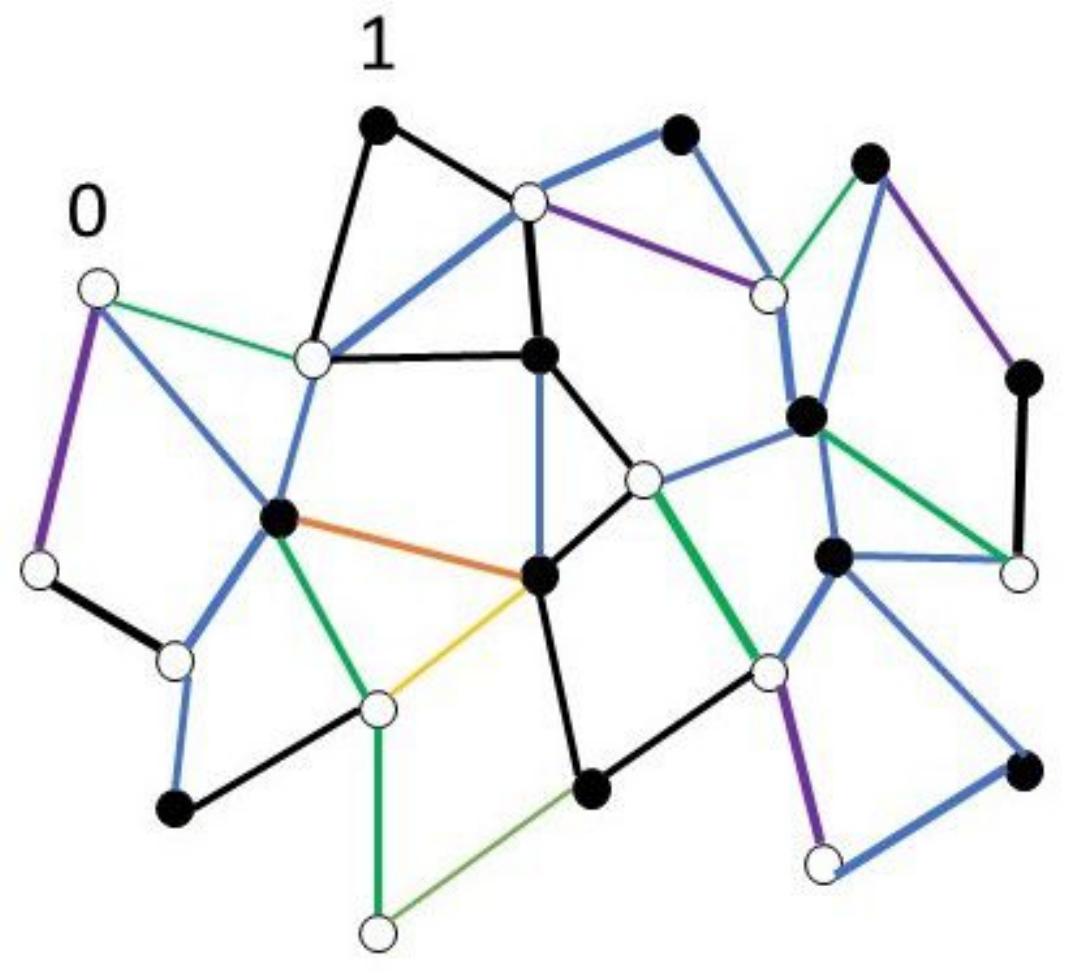}
  \caption{The lattice model is the dispersion of continuous space, and a typical quantum simulation is the simulation of lattice model. The correlation between each discrete space point and the state of different points is encoded into the quantum state of the quantum simulator.}\label{fig2}
  \end{center}
  \end{figure}
Nowadays, with the developing of quantum technology, quantum computer shows a bright prospects. Once the key to connect quantum information with various fields found. We can use the powerful of quantum computers to solve the problems. This key is the universal password alone with the evolution of us, that is logic. In modern mathematics, the language of logic is category theory. Of course, the quantum circuit itself form a structure of category. If we find the relation between the category of problem and the category of quantum circuit, then we can use the quantum computer to solve the problem.

In this paper, we propose a new paradigm for quantum simulation based on the mathematical tool of category theory. The new paradigm overcame the shortcomings of the Feynman paradigm and greatly expanded the scope of quantum simulation. In addition, our quantum simulation paradigm introduces a different coding method than the original, which was based on lattice model, which is completely geometrical, so its coding is a reductionist idea, whereas we use the mathematical structure of category theory. It is highly abstract and highly encapsulated, highly distilling information from simulated systems and turning it into something algebraic. So our quantum simulation paradigm introduces a new way of coding. It is not reductionism but an emergenism.

This paper is organized as follows. First, starting with a very brief introduction to category of quantum circuit and other systems. Next, we will  define quantum simulators in category language. Then we will briefly introduce how the tensor category describes other systems and give a definition of quantum simulation that goes beyond Feynman's paradigm. Finally, this paper will demonstrate categorized quantum simulation with simulating the Yang-Mills theory as an example.

\begin{figure}[h]
\begin{center}
\includegraphics[width=\textwidth]{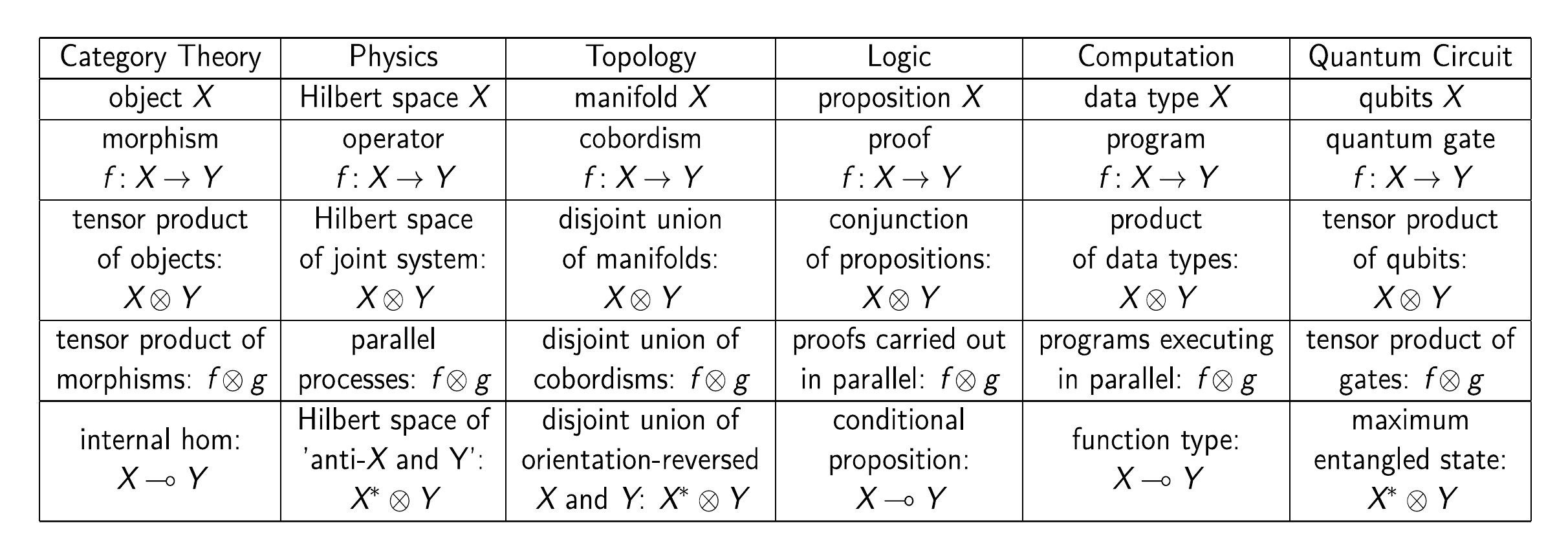}
\caption{Categorical description of physics, topology, logic, computer science, quantum circuits}\label{catt}
\end{center}
 \end{figure}

\begin{figure}[h]
\begin{center}
\includegraphics[width=0.6\textwidth]{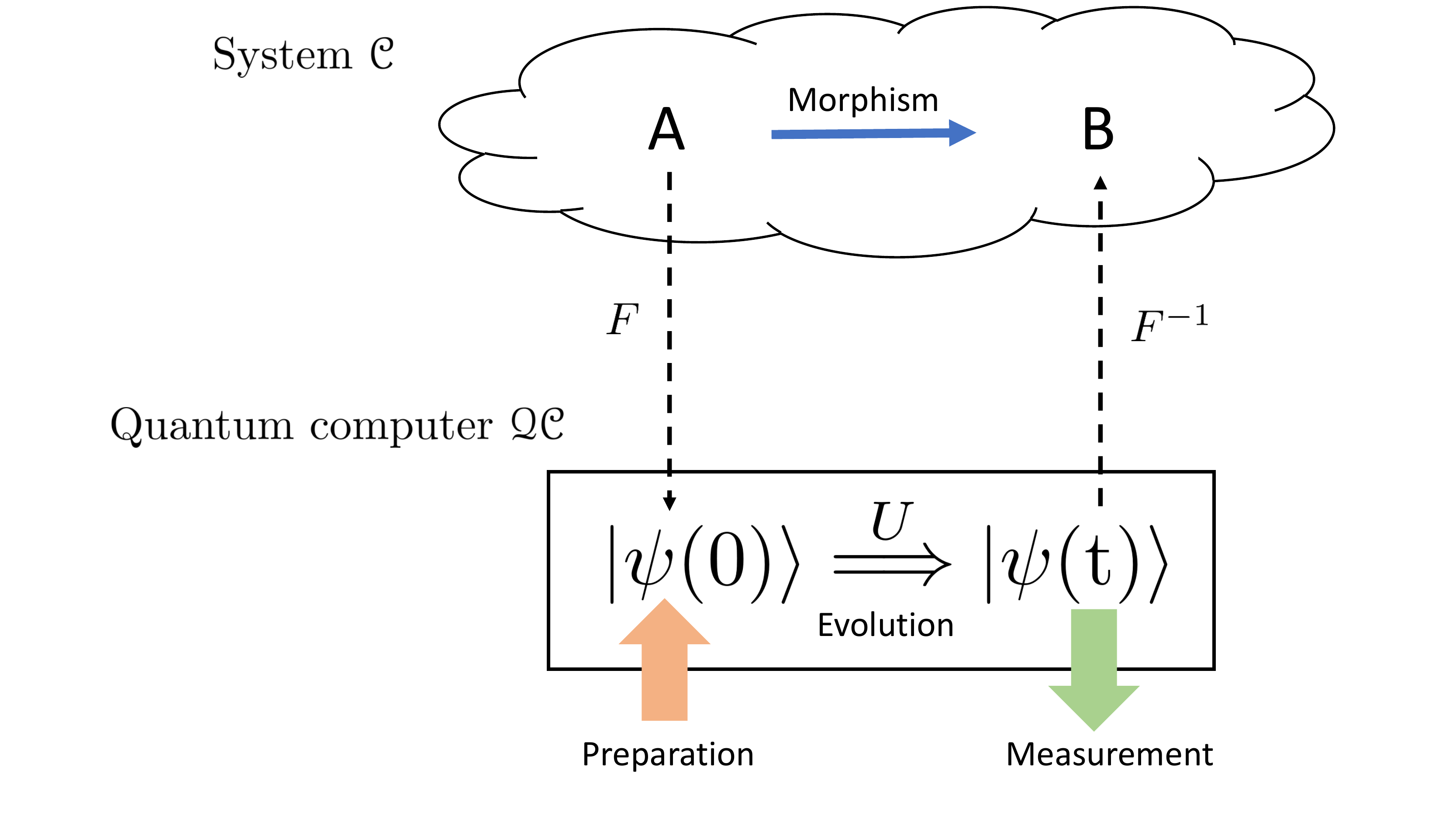}
\caption{Categorical quantum simulation}\label{cqs}
\end{center}
\end{figure}

\paragraph{The category of quantum circuit $\mathbf{QC}$}
Article\cite{Bergholm2011CategoricalQC} gives the category structure of quantum circuit, it is a $\dagger$-compact catgory, that consists of
\begin{itemize}
\item[1.] Object $A:=(\EuScript{A},D_A)$, where $D_A = (d_{A_i})^{n_{A}}_{i=1}$is the dimension of sub-spac, $\EuScript{A} = \mathbb{C}_{d_1} \otimes \mathbb{C}_{d_2}\otimes ... \otimes \mathbb{C}_{d_{n_A}}$are finite dimension Hilbert space. The dimension of $A$ is $\mathrm{dim} A := \mathrm{dim}\EuScript{A} = \prod_{i=1}^{n_{A}} d_{A i}$, where $n_A$ represents the dimension of sub-space of object. If $n_A=1$ it is called simple object, otherwise called composite object. For the Hilbert space in each object $A$ we shall choose a computational basis (equal to the standard
tensor basis), denoted by $\left\{\left|i_{1} i_{2} \ldots i_{n_{A}}\right\rangle_{A}\right\}_{i_{k}=0}^{d_{A k}-1}$. 
\item[2.]  For any object $A$, $B$. The set of morphism $\mathrm{Hom}_{\mathbf{QC}}(A,B)$contains all the bounded linear maps from $A$ to $B$. $D_A$ and $D_B$ are the dimension of the input and output of the bounded linear maps.  Unitary morphisms of $\mathbf{QC}$ are the quantum gates.

\item  [3.] Composites of morphisms $\circ$is the composite of linear maps.
\item  [4.] The definitions of  tensor product functor $\otimes$ and identity $
\mathbf{1}:=(\mathbb{C}, (1))$ as follow: 

The tensor product of objects: $A \otimes B:=(A\otimes B, D_A \star D_B)$, where $\star$ is the dimension of the composed space 

The tensor product of morphisms: $f  \otimes g$, where $f:A \rightarrow A'$, $g:B \rightarrow B'$,is given
by the Kronecker product of the corresponding matrices in the computational basis:
\begin{equation}
\left\langle\left. i j\right|_{A^{\prime} \otimes B^{\prime}}(f \otimes g) \mid p q\right\rangle_{A \otimes B}:=\left\langle\left. i\right|_{A^{\prime}} f \mid p\right\rangle_{A}\left\langle\left. j\right|_{B^{\prime}} g \mid q\right\rangle_{B}
\end{equation}
\item[5.] $\dagger$ functor, which is identity on the objects and takes the Hermitian adjoint of the morphisms
\item[6.] For every object A the unit and counit morphisms, defined in terms of the computational basis:
\begin{equation}
\eta_{A}:=\sum_{k}|k k\rangle_{A \otimes A}, \quad \epsilon_{A}:=\sum_{k} \langle k k|_{A \otimes A}=\eta_{A}^{\dagger}.
\end{equation}
Every object is its own dual:$A^{*}=A$.
\end{itemize}
\paragraph{Categorical description of physics, topology, logic and computer science}
Tensor category descriptions for each discipline are arranged in the following diagram~\ref{catt}.\cite{Baez2009PhysicsTL}. Figure~\ref{catt}. is like a multilingual  dictionary, in which unknown fields can be translated directly into familiar subjects with a clear understanding of a subject. These disciplines can also be translated into quantum circuits. This provides a solid foundation for quantum simulations.
\paragraph{Definition of categorical quantum simulation}
Here we gives the definition of categorical quantum simulation: categorical quantum simulation is a functor 
$F$ maps the category of simulated system $\EuScript{C}$ to the category of quantum circuit$\mathbf{QC}$
\begin{equation}
F:\EuScript{C}\rightarrow\mathbf{QC}
\end{equation}
At the same time, the inverse functor $F^{-1}$ exist, it  maps the quantum circuit category $\ mathbf {QC} $ back to the category of simulated system$\EuScript{C}$:
\begin{equation}
F^{-1}:\mathbf{QC}\rightarrow \EuScript{C}
\end{equation}
See Figure~\ref{cqs} for a specific exchange diagram. In categorized quantum simulations, it is no longer required that the simulated system must have a swarm algebraic structure, as long as it has objects and states of emission, so almost all systems can meet this condition. Secondly, the strong encapsulation ability based on tensor domain does not require lattice model to encode. Simply encoding the object directly saves quantum bit resources. This is  the emergenism coding.
\paragraph{Topological quantum field theory }
Topological quantum field theory is defined by Atiyah and Segal as a functor between two special categories\cite{LIHS2003TopologicalQF,Segal1988TheDO}. One is a geometric category and the other is an algebraic category. The geometric category is the category of bordisms denote as $\mathbf{Bord}_{d}$. For every non-negative integer $d$ there is a category corresponding to the $d$ dimensional topological field theory. The object of $d$ dimension bordisms category is a $d-1$ dimension manifold, and the morphism   is a $d$ dimension bordisms between the manifolds. The algebraic category is the vector space category over  $\mathbb{C}$ denote as $\mathbf{Vect}_{\mathbb{C}}$. Topological field theory is defined as functor maps geometric category $\mathbf{Bord}_{d}$ to the algebraic category $\mathbf{Vect}_{\mathbb{C}}$, denote as
\begin{equation}
Z: \mathbf{Bord}_{d} \rightarrow \mathbf{Vect}_{\mathbb{C}}
\end{equation}
Both categories have additional structures that make them both symmetric monoidal categories. Therefore, the corresponding functor of the quantum topological field theory is a symmetric monoidal functor $Z$. The symmetric monoidal structure of bordisms  category is given by union of manifolds. The symmetric monoidal structure of algebraic category is given by the tensor product of vector space. So topological field theory combines geometry and generation by functor $Z$\cite{Dijkgraaf1989AGA,Voronov1994TopologicalFT,StreetBaltimore1996TwodimensionalTQ}.But even more surprising is that topological field theory explains a new algebraic structure\cite{Sawin1995DirectSD,Quinn1991LecturesOA,Dubrovin1994GeometryO2,Kock2004FrobeniusAA}. Folklore gave the theorem that a two-dimensional bordism category is a free-symmetric monoidal category, and its object has the structure of an commutative Frobenius algebra. In particular, this two-dimensional bordism category is equivalent to exchanging commutative $k$-algebraic category.

\begin{figure}[h]
  \begin{center}
 \includegraphics[width=0.9\textwidth]{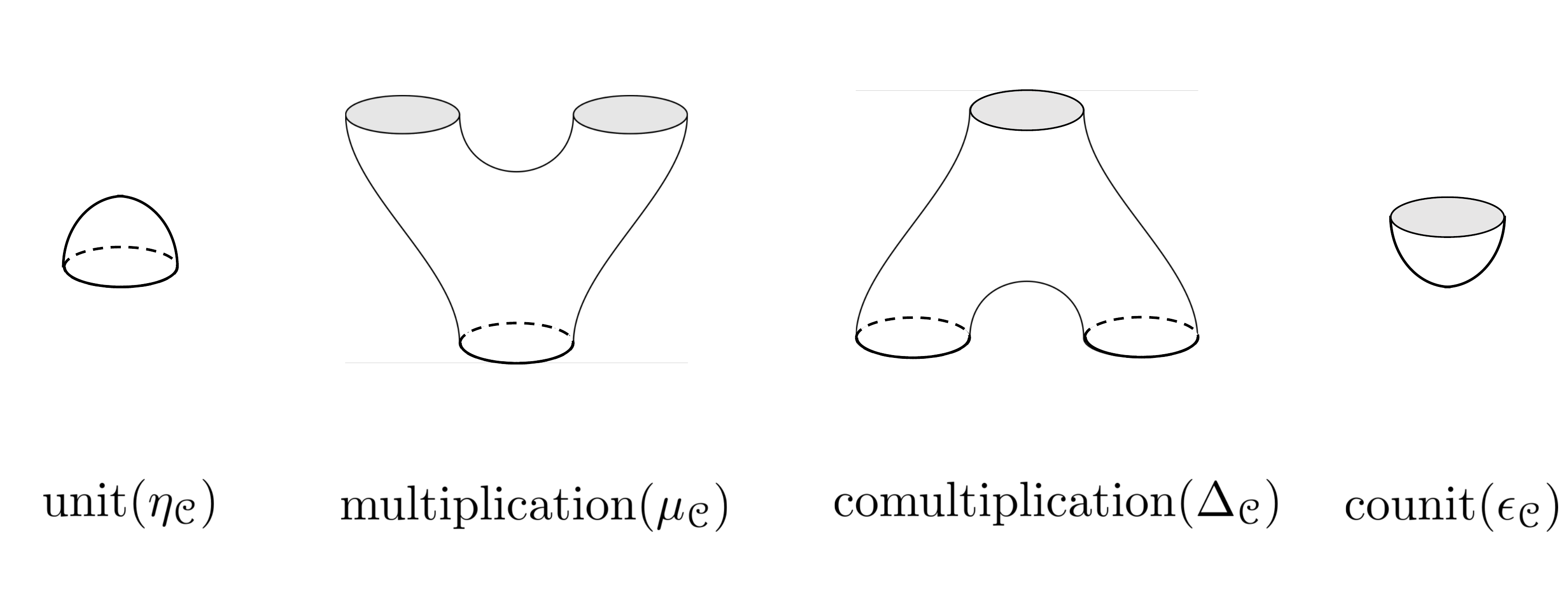}
  \caption{wo-dimensional is a commutative Frobenius algebra}\label{cfa}
  \end{center}
  \end{figure}

Folklore's theorem indicates a two-dimensional bordism category is equivalent to the symmetric monoidal category, generated by a simple object $S^{1}$ with morphisms shown in Figure~\ref{cfa}.
The theorem also indicates that a 2-dimensional topological quantum field theory is determined entirely by the vector space $Z(S^{1})$ and the linear mapping corresponding to each object generated by the application of topological quantum field operations. These linear maps are:
\begin{equation}
\begin{aligned}
\eta_{\EuScript{C}}:Z(\phi)=\mathbb{C} &\rightarrow Z(S^{1})\\
\mu_{\EuScript{C}}: Z(S^{1})\otimes Z(S^{1}) &\rightarrow Z(S^{1})\\
\Delta_{\EuScript{C}}: Z(S^{1})&\rightarrow Z(S^{1})\otimes Z(S^{1})\\
\epsilon_{\EuScript{C}}: Z(S^{1}) &\rightarrow Z(\phi)=\mathbb{C}
\end{aligned}\label{cfaa}
\end{equation}  
where $\phi$ is the  empty set.
The Frobenius algebra corresponding to the 2d-Yang-Mills field is a special case of topological field theory\cite{Donnelly2019EntanglementBM}:
\begin{figure}[h]
\centering
\includegraphics[width=0.6\textwidth]{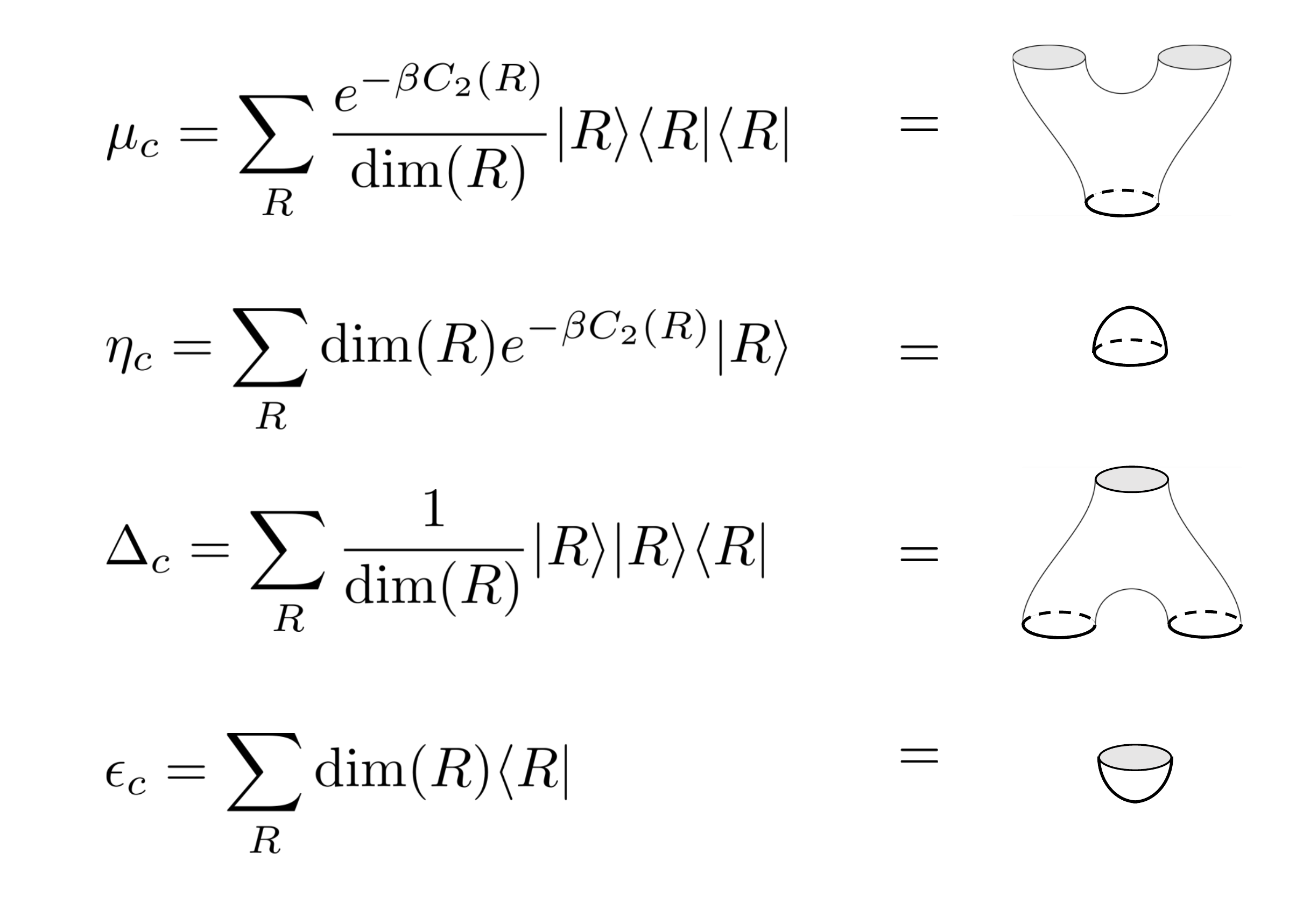}
\caption{Frobenius algebra of 2d-Yang-Mills field }\label{ymmmm}
\end{figure}
The visible Figure~\ref {ymmmm} maps the bordism category to vector space, and if the bordism category is mapped to the quantum circuit category, the categorical quantum simulation can be completed. Since the operator in Figure~\ref{ymmmm} is a non-unitary evolution operator, its implementation in a quantum circuit requires the duality mode of quantum computing\cite{Guilu2006GeneralQI,Long2008DualityQC,Guilu2009AllowableGQ}.

\begin{figure}[h]
\centering
\includegraphics[width=\textwidth]{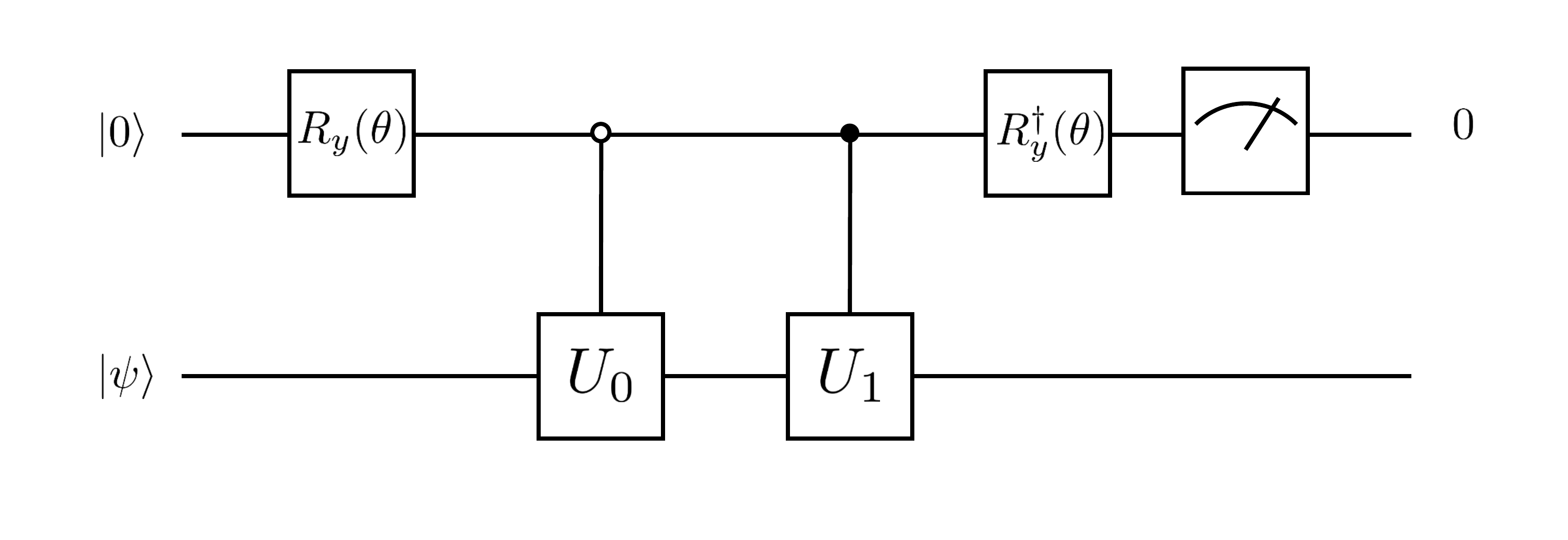}
\caption{Quantum circuit of duality quantum computing for an auxiliary bit and a working bit}\label{1dqc}
\end{figure}

\begin{figure}[h]
\centering
\includegraphics[width=0.9\textwidth]{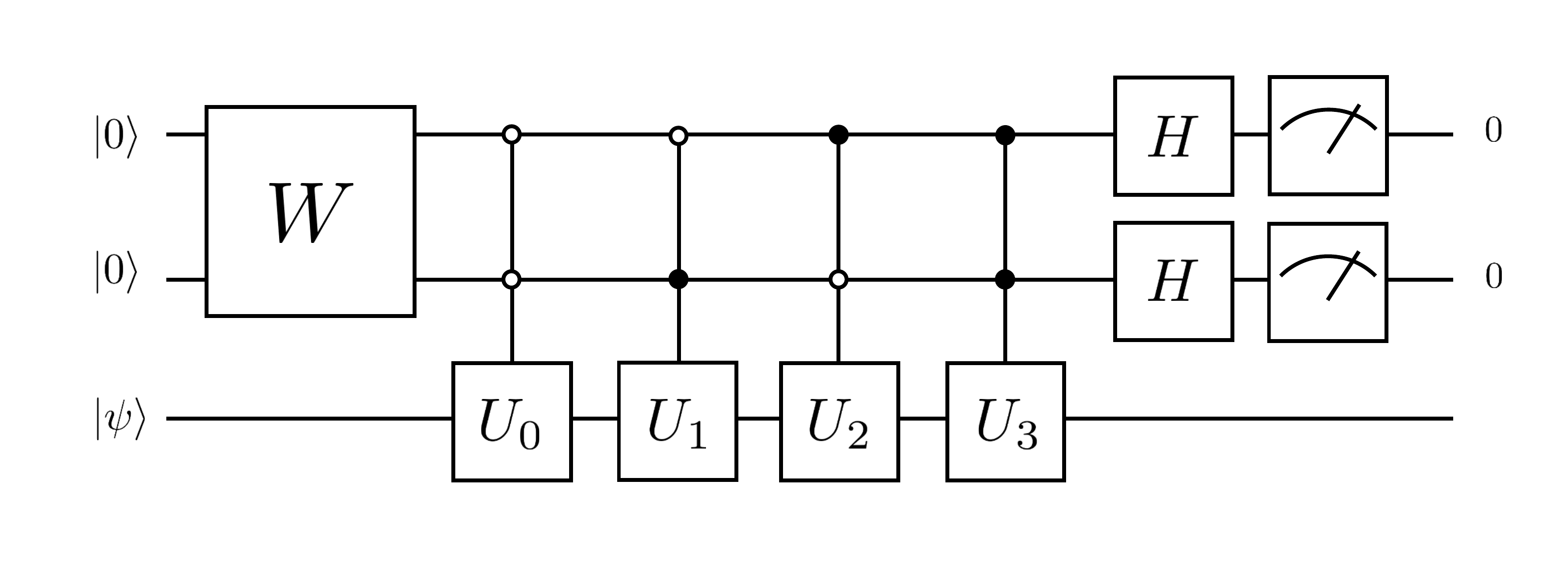}
\caption{Quantum circuit of duality quantum computing for two auxiliary bits and one working bit}~\label{2dqc}
\end{figure}

\paragraph{The Categorical quantum simulation of $SU(3)$ Yang-Mills Theory}
To simplify the model, here we do a truncation where the $R$ only traverses the first three irreducible  representations of $SU(3)$ gauge group. According to the group representation theory of $SU (3)$. The eigenvalue of Casimir operator $\hat{C}_2$ is:
\begin{equation}
 \frac{4}{3}(p^2+q^2+pq+3p+3q)
\end{equation}
the dimension of which is
\begin{equation}
\mathrm{dim}D(p,q)=\frac{1}{2}(p+1)(q+1)(p+q+2)
\end{equation}
for $D(0,0)$ representation
\begin{equation}
 C_{2}\left(D(0,0)\right)=\frac{4}{3}(0^2+0^2+0+3\times 0+3\times 0)=0
\end{equation}
\begin{equation}
\mathrm{dim}\left(D(0,0)\right)=\frac{1}{2}(0+1)(0+1)(0+0+2)=1
\end{equation}
for $D(1,0)$ representation
\begin{equation}
 C_{2}\left(D(1,0)\right)=\frac{4}{3}(1^2+0^2+1\times 0+3\times 1+3\times 0)=\frac{16}{3}
\end{equation}
\begin{equation}
\mathrm{dim}\left(D(1,0)\right)=\frac{1}{2}(1+1)(0+1)(1+0+2)=3
\end{equation}
for $D(0,1)$ representation
\begin{equation}
 C_{2}\left(D(0,1)\right)=\frac{4}{3}(0^2+1^2+0\times 1+3\times 0+3\times 1)=\frac{16}{3}
\end{equation}
\begin{equation}
\mathrm{dim}\left(D(0,1)\right)=\frac{1}{2}(0+1)(1+1)(0+1+2)=3
\end{equation}
Based on category theory it can be coded as follows
\begin{figure}[h]
\centering
\includegraphics[width=0.7\textwidth]{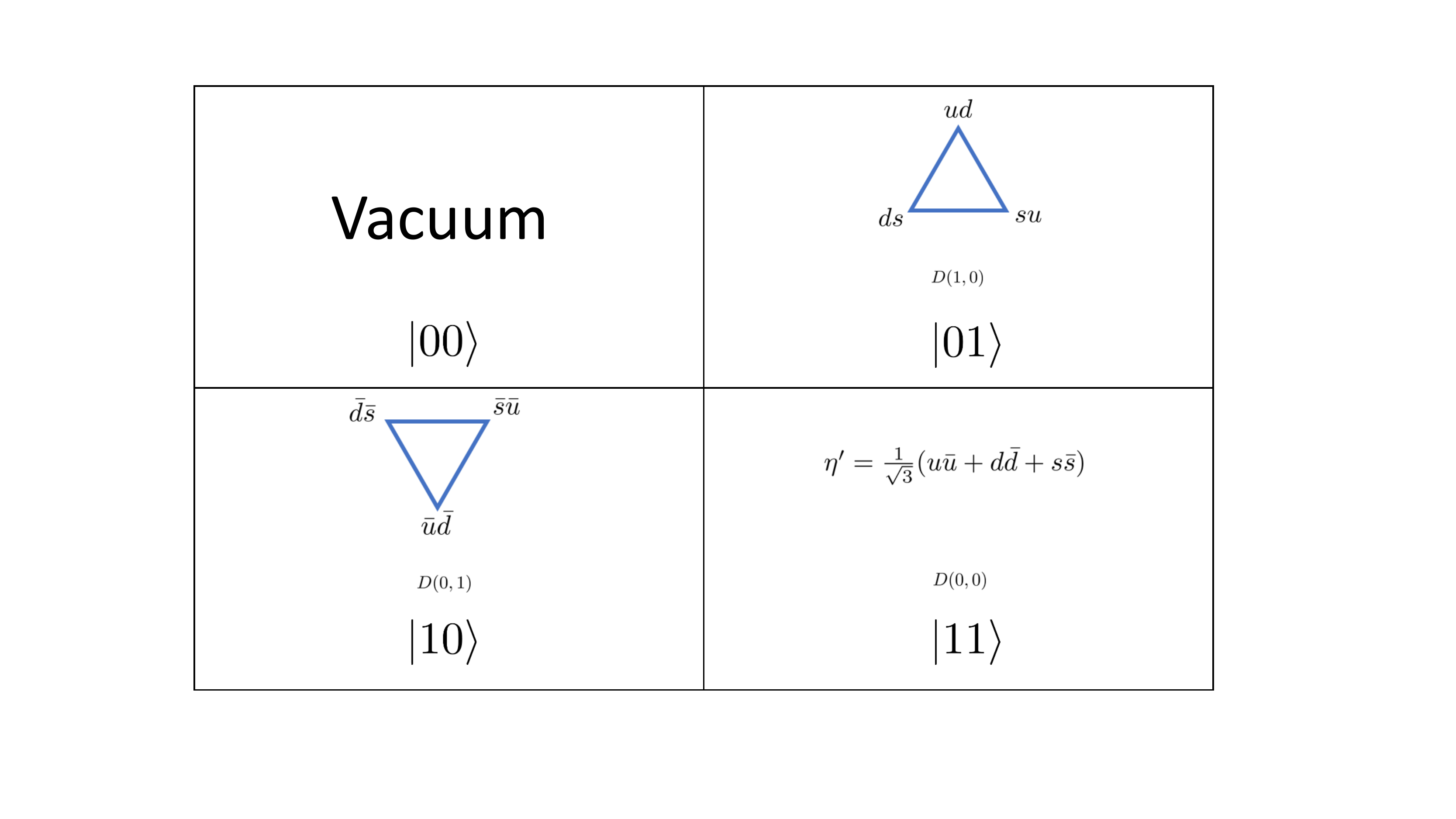}
\caption{The coding of categorical quantum  Simulation of $SU(3) $ Yang-Mills Theory}
\label{bm}
\end{figure}
Folklore's theorem shows that the Frobenius algebra shown in Figure~\ref{ymmmm} is the 2d Yang-Mills field theory. Mapping Frobenius algebra onto quantum circuit enables categorical quantum simulation of the $SU(3)$ Yang-Mills theory. This process is a functor:
\begin{equation}
F:\mathbf{Bord}_{2}\rightarrow \mathbf{QC}\label{fclzmn}
\end{equation}
Map the bordism category $\ mathbf {Bord} _ {2} $ to the quantum circuit category $\ mathbf {QC} $. This part  will be completed next.
For the first operator of Frobenius
\begin{equation}
\mu_{c}=\sum_{R} \frac{e^{-\beta C_{2}(R)}}{\operatorname{dim}(R)}|R\rangle\langle R|\langle R|\label{536}
\end{equation}
To ensure that the quantum circuit is reversible, a vacuum state in the incident state of the formula Eq.(~\ref{536}) is written.
\begin{equation}
\mu_{c}=\sum_{R} \frac{e^{-\beta C_{2}(R)}}{\operatorname{dim}(R)}|R\rangle|0\rangle\langle R|\langle R|\label{537}
\end{equation}
After coding, take the natural unit system, that is, $\ beta = $1, then formula ~\ref{537} can be written
\begin{equation}
\mu_{c}=|1100\rangle\langle1111|+\frac{1}{3}e^{-\frac{16}{3}i}|1000\rangle\langle1010|+\frac{1}{3}e^{-\frac{16}{3}i}|0100\rangle\langle0101|\label{538}
\end{equation}
s you can see, the formula Eq.(\ref {538}) is not unitary, but non-unitary. So if you want to do this with quantum circuits, you have to do it with the duality model of quantum computing. First disassemble the formula Eq.(~\ref{538}) quantum states into the form of a direct product.
\begin{equation}
\begin{split}
|1\rangle\langle1|\otimes |1\rangle\langle1| \otimes |0\rangle\langle1| \otimes |0\rangle\langle1|
+&\frac{1}{3}e^{-\frac{16}{3}i}( |1\rangle\langle1|\otimes|0\rangle\langle0|\otimes|0\rangle\langle1|\otimes|0\rangle\langle0|\\
+&|0\rangle\langle0|\otimes|1\rangle\langle1|\otimes|0\rangle\langle0|\otimes|0\rangle\langle1| )
\end{split}
\end{equation}
Using the Pauli operator base to expand the above
\begin{equation}
\begin{split}
\left(0.5\sigma_{0}-0.5\sigma_{z}\right)\otimes\left(0.5\sigma_{0}-0.5\sigma_{z}\right)\otimes\left(0.5\sigma_{x}+0.5i\sigma_{y}\right)\otimes\left(0.5\sigma_{x}+0.5i\sigma_{y}\right)
\\
+\frac{1}{3}e^{-\frac{16}{3}i}\left[\left(0.5\sigma_{0}-0.5\sigma_{z}\right)\otimes\left(0.5\sigma_{0}+0.5\sigma_{z}\right)\otimes\left(0.5\sigma_{x}+0.5i\sigma_{y}\right)\otimes\left(0.5\sigma_{0}+0.5\sigma_{z}\right)\right]
\\
+\frac{1}{3}e^{-\frac{16}{3}i}\left[\left(0.5\sigma_{0}+0.5\sigma_{z}\right)\otimes\left(0.5\sigma_{0}-0.5\sigma_{z}\right)\otimes\left(0.5\sigma_{0}+0.5\sigma_{z}\right)\otimes\left(0.5\sigma_{x}+0.5i\sigma_{y}\right)\right]
\end{split}
\end{equation}
Tidying up the upper equations in the form of Pauli matrix coefficients
\begin{equation}\
\begin{split}
\left[\left(0.5+\frac{1}{3}e^{-\frac{16}{3}i}\right)\sigma_{0}-0.5\sigma_{z}\right]&\otimes\left[1.5\sigma_{0}-0.5\sigma_{z}\right]\\
&\otimes \left[0.5\sigma_{0}+\sigma_{x}+i\sigma_{y}+0.5\sigma_{z}\right]^{\otimes 2}  
\end{split}
\end{equation}
normalized and get
\begin{equation}
\begin{split}
\left[0.598e^{-0.372i}\sigma_{0}-0.402\sigma_{z}\right]\otimes &\left[0.75\sigma_{0}-0.25\sigma_{z}\right]\\
\otimes &\left[\frac{1}{\sqrt{10}}\sigma_{0}+\frac{2}{\sqrt{10}}\sigma_{x}+\frac{2}{\sqrt{10}}i\sigma_{y}+\frac{1}{\sqrt{10}}\sigma_{z}\right]^{\otimes 2}
\end{split}\label{542}
\end{equation}
In order to apply the above formula to the duality mode of quantum computing. Next we need to determine the $W$ and $V$ operators. Here we first calculate the general form of duality quantum computing for an auxiliary bit and a working bit (Figure~\ref{1dqc}).Let $V=R_{y}(\theta)$, $W=R^{\dagger}_{y}(\theta)$
where
\begin{equation}
R_{y}(\theta)=e^{-\frac{i \sigma_{y}\theta}{2}}=\left(\begin{array}{cc}
\cos\left(\frac{\theta}{2}\right) & -\sin\left(\frac{\theta}{2}\right) \\
\sin\left(\frac{\theta}{2}\right) & \cos\left(\frac{\theta}{2}\right)
\end{array}\right)
\end{equation}
The initial state evolves after $R_{y}(\theta)$ with two controlled operations of $U_{0}$ and $U_{1}$
\begin{equation}
\cos\left(\frac{\theta}{2}\right)|0\rangle U_{0} |\psi\rangle+\sin\left(\frac{\theta}{2}\right)|1\rangle U_{1} |\psi\rangle
\end{equation}
Evolved after $R^{\ dagger}_{y}(\theta)$
\begin{equation}
\begin{split}
\cos^{2}\left(\frac{\theta}{2}\right)|0\rangle U_{0} |\psi\rangle-&\cos\left(\frac{\theta}{2}\right)\sin\left(\frac{\theta}{2}\right)|1\rangle U_{0} |\psi\rangle\\
+&\sin^{2}\left(\frac{\theta}{2}\right)|0\rangle U_{1} |\psi\rangle+\cos\left(\frac{\theta}{2}\right)\sin\left(\frac{\theta}{2}\right)|1\rangle U_{1} |\psi\rangle
\end{split}
\end{equation}
If the auxiliary bit is measured in state 0, the final state of working bit is
\begin{equation}
\left[\cos^{2}\left(\frac{\theta}{2}\right) U_{0}+\sin^{2}\left(\frac{\theta}{2}\right) U_{1}\right] |\psi\rangle \label{546}
\end{equation}
Next we calculate the general form of duality quantum computing for two auxiliary bits and one working bit (see Figure~\ref{2dqc}).
The initial state evolves by the function of the constructed operator $W$
\begin{equation}
\{c_{0}|00\rangle+c_{1}|01\rangle+c_{2}|02\rangle+c_{3}|03\rangle\}|\psi \rangle\label{547}
\end{equation}
After passing through  controlled quantum gates $U_{0}$, $U_{1}$, $U_{2}$, $U_{3}$, the quantum states becomes
\begin{equation}
c_{0}|00\rangle U_{0}|\psi \rangle+c_{1}|01\rangle U_{1}|\psi \rangle+c_{2}|10\rangle U_{2}|\psi \rangle+c_{3}|11\rangle\ U_{3}|\psi \rangle
\end{equation}
After  $H$ transformation, the final state evolves to
\begin{equation}
\begin{split}
0.25&(|00\rangle+|01\rangle+|10\rangle+|11\rangle) c_{0}U_{0}|\psi \rangle+0.25(|00\rangle-|01\rangle+|10\rangle-|11\rangle)c_{1} U_{1}|\psi \rangle\\
+&0.25(|00\rangle+|01\rangle-|10\rangle-|11\rangle) c_{2} U_{2}|\psi \rangle+0.25(|00\rangle-|01\rangle-|10\rangle+|11\rangle) c_{3}U_{3}|\psi \rangle
\end{split}
\end{equation}
If the auxiliary bit is measured in state 0, the final state of working bit is
\begin{equation}
\left[c_{0}U_{0}+c_{1}U_{1}+c_{2}U_{2}+c_{3}U_{3}\right]|\psi\rangle
\end{equation}

\begin{figure}[h]
\centering
\includegraphics[width=0.7\textwidth]{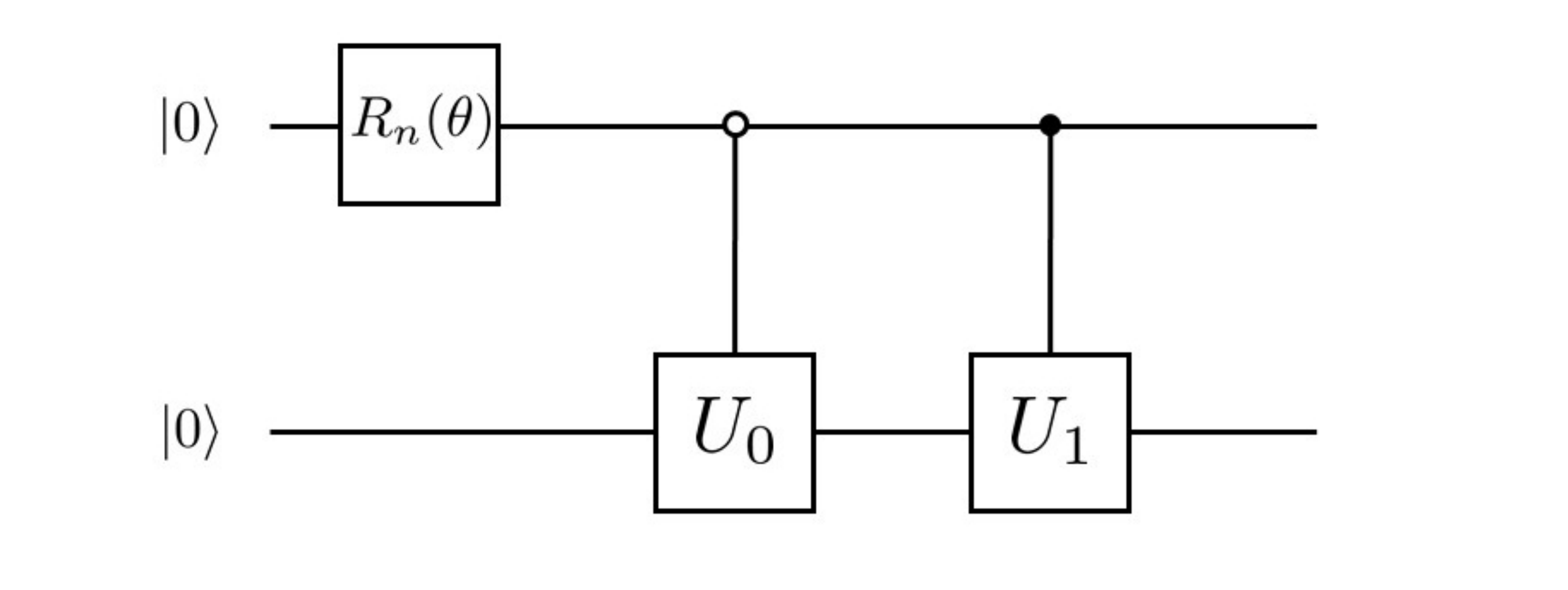}
\caption{Preparation of two-bit arbitrary quantum states}\label{aws}
\end{figure}

The following will discuss the construction of the operator $W$ in Eq.(~\ref{547}). The effect of the operator $W$ is shown in Figure ~\ref{aws}.
$R_{n}(\theta)$ gate can be constructed first, making
\begin{equation}
|00\rangle \rightarrow \left(\sqrt{c^{2}_{0}+c^{2}_{1}}|0\rangle+\sqrt{c^{2}_{2}+c^{2}_{3}}|1\rangle\right)|0\rangle\label{551}
\end{equation}
he second step is to construct a controlled quantum gate of $U_{1}$ and $U_{2}$ so that the above evolves to
\begin{equation}
\begin{split}
\sqrt{c^{2}_{0}+c^{2}_{1}}&|0\rangle \left(\frac{c_{0}}{\sqrt{c^{2}_{0}+c^{2}_{1}}}|0\rangle+\frac{c_{1}}{\sqrt{c^{2}_{0}+c^{2}_{1}}}|1\rangle\right)\\
+&\sqrt{c^{2}_{2}+c^{2}_{3}}|1\rangle \left(\frac{c_{2}}{\sqrt{c^{2}_{2}+c^{2}_{3}}}|0\rangle+\frac{c_{3}}{\sqrt{c^{2}_{2}+c^{2}_{3}}}|1\rangle\right)
\end{split}\label{552}
\end{equation}
where
\begin{equation}
\begin{split}
U_{1}|0\rangle &\rightarrow \frac{c_{0}}{\sqrt{c^{2}_{0}+c^{2}_{1}}}|0\rangle+\frac{c_{1}}{\sqrt{c^{2}_{0}+c^{2}_{1}}}|1\rangle\\
U_{2}|0\rangle &\rightarrow \frac{c_{2}}{\sqrt{c^{2}_{2}+c^{2}_{3}}}|0\rangle+\frac{c_{3}}{\sqrt{c^{2}_{2}+c^{2}_{3}}}|1\rangle
\end{split}\label{553}
\end{equation}

The following is a quantum circuit implementation of the Eq.(~\ref{542}). First, look at the fact that the first qubit has two operations and therefore requires the introduction of an auxiliary bit. control Eq.(~\ref{546}). Make the phase gate $e^{-0.372 i}\sigma_{0}=U _ {0}$, $-\sigma_{z}=U_{1}$, then for the operator $R_{y}(\theta_{1})$ can be determined by the Eq.(~\ref{546}) its angle is $\theta_{1}$
\begin{equation}
\cos^{2}\left(\frac{\theta_{1}}{2}\right)=0.598
\end{equation}
and get$\theta_{1}$=1.37

Then there are the two operations of the second qubit that require the introduction of an auxiliary qubit. compare with Eq.(~\ref{546}). Make the phase gate $\sigma_{0}=U_{0}$, $\sigma_{z}=U_{1}$, then for the operator $R_{y}(\theta_{2})$ can be determined by the Eq.(~\ref{546}) its angle $\ theta _ {2} $,
\begin{equation}
\cos^{2}\left(\frac{\theta_{2}}{2}\right)=0.75
\end{equation}
get $\theta_{2}=\frac{\pi}{3}$

See also that the third qubit has three operations and therefore requires the introduction of two auxiliary qubits to implement the $c_{0}U_{0}+c_{1}U_{1}+c_{2}U_{2}+c_{3}U_{3}$. Make the phase gate $\sigma_{0}=U_{0}$, $\sigma_{x}=U_{1}$, $i\sigma_{y}=U_{2}$, $\sigma_{z}=U_{3}$. Constructing the gate $R_{n}(\theta)$ according to the Eq.(~\ref{551})
\begin{equation}
R_{n}(\theta)|00\rangle \rightarrow \left(\frac{\sqrt{2}}{2}|0\rangle+\frac{\sqrt{2}}{2}|1\rangle\right)|0\rangle\
\end{equation}
get $R_{n}(\theta)$ as $R_{y}(\frac{\pi}{2})$

\begin{figure}[h]
\centering
\includegraphics[width=0.9\textwidth]{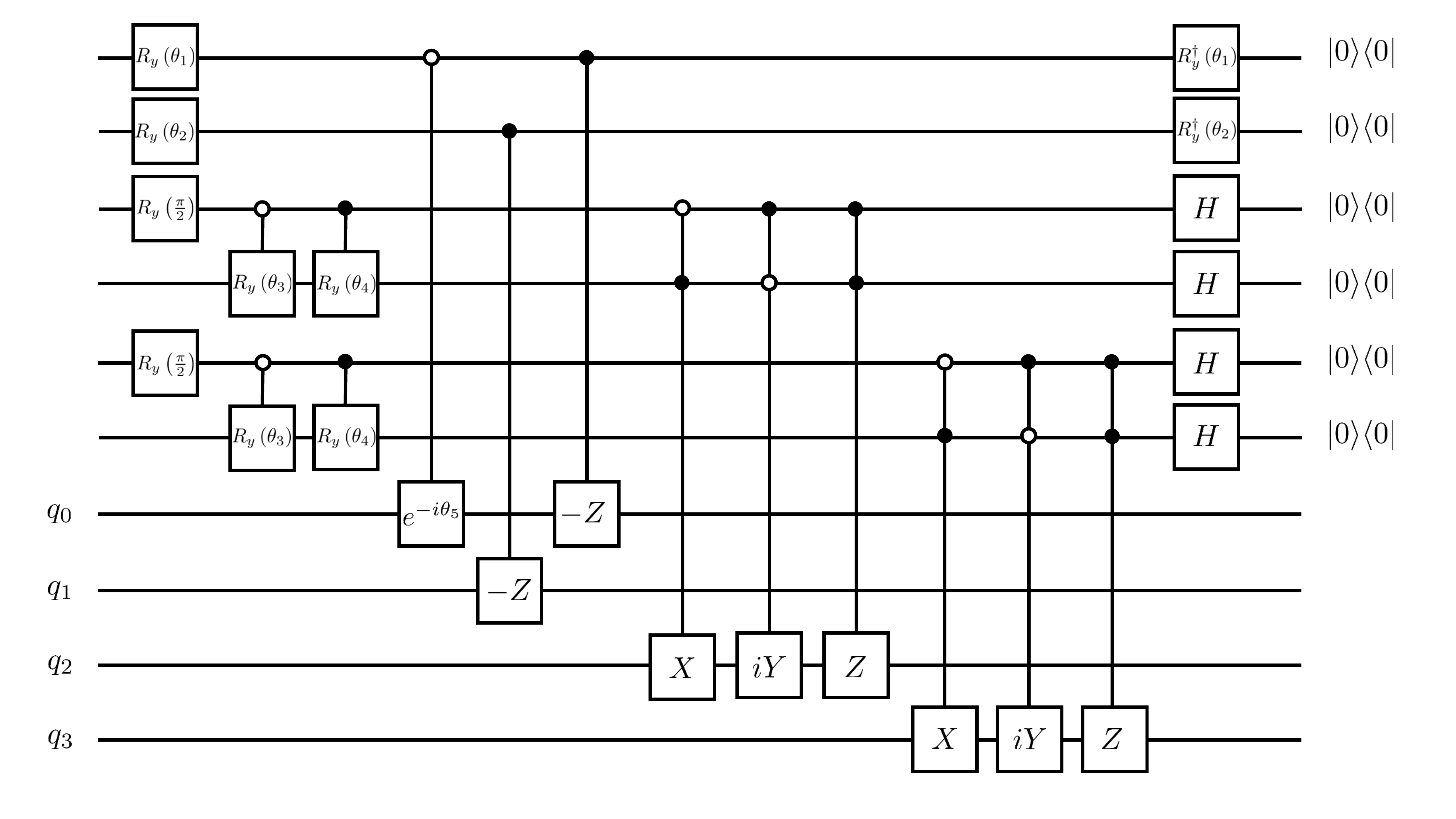}
\caption{quantum circuit of the operator $\mu_{c}$,where $q_{0}, q_{1}, q_{2}, q_{3}$are four working qubits, $\theta_{1}=1.37$, $\theta_{2}=\frac{\pi}{3}$ ,$\theta_{3}=2.21$, $\theta_{4}=0.93$, $\theta_{5}=0.372$ }\label{ym01}
\end{figure}

According to the Eq.(~\ref{553}) two controlled operators $U_{1}$  and $U_{2}$ can be obtained to:
\begin{equation}
\begin{split}
U_{1}|0\rangle &\rightarrow \frac{\sqrt{5}}{5}|0\rangle+\frac{2\sqrt{5}}{5}|1\rangle\\
U_{2}|0\rangle &\rightarrow \frac{2\sqrt{5}}{5}|0\rangle+\frac{\sqrt{5}}{5}|1\rangle
\end{split}
\end{equation}
get $U_{1}=R_{y}(\theta_{3})$, $U_{2}=R_{y}(\theta_{4})$, where$\theta_{3}=2.21$, $\theta_{4}=0.93$
For the fourth working bit and the third working bit are the same. The quantum circuit of the operator $\mu_{c}$ can be shown as follows Figure~\ref{ym01}

The second part of Yang-Mills field theory is disassembled below.
\begin{equation}
\Delta_{c}=\sum_{R} \frac{1}{\operatorname{dim}(R)}|R\rangle |R\rangle \langle R|
\end{equation}
In order to ensure that the quantum circuit is reversible, a vacuum state is added to the upper ejection state.
\begin{equation}
\Delta_{c}=\sum_{R} \frac{1}{\operatorname{dim}(R)}|R\rangle |R\rangle \langle R|\langle 0|
\end{equation}
After coding, take the natural unit system, that is, $\beta=1$, and the above can be written
\begin{equation}
\Delta_{c}=|1111\rangle\langle1100|+\frac{1}{3}|1010\rangle\langle1000|+\frac{1}{3}|0101\rangle\langle0100|
\end{equation}
First, the superposed quantum states are disassembled into direct product forms
\begin{equation}
\begin{split}
|1\rangle\langle1|\otimes |1\rangle\langle1| \otimes |1\rangle\langle0| \otimes |1\rangle\langle0|
+&\frac{1}{3}( |1\rangle\langle1|\otimes|0\rangle\langle 0 |\otimes|1\rangle\langle 0|\otimes|0\rangle\langle0|\\
+&|0\rangle\langle0|\otimes|1\rangle\langle1|\otimes|0\rangle\langle0|\otimes|1\rangle\langle0| )
\end{split}
\end{equation}
Expand with Pauli substrate available
\begin{equation}
\begin{split}
\left(\sigma_{0}-\sigma_{z}\right)\otimes& \left(\sigma_{0}-\sigma_{z}\right) \otimes \left(\sigma_{x}-i\sigma_{y}\right) \otimes \left(\sigma_{x}-i\sigma_{y}\right)\\
+& \left(\frac{1}{3}\sigma_{0}-\frac{1}{3}\sigma_{z}\right)\otimes\left(\sigma_{0}+\sigma_{z}\right)\otimes\left(\sigma_{x}-i\sigma_{y}\right)\otimes\left(\sigma_{0}+\sigma_{z}\right)\\
+&\left(\frac{1}{3}\sigma_{0}+\frac{1}{3}\sigma_{z}\right)\otimes\left(\sigma_{0}-\sigma_{z}\right)\otimes\left(\sigma_{0}+\sigma_{z}\right)\otimes\left(\sigma_{x}-i\sigma_{y}\right) 
\end{split}
\end{equation}
Tidying up the upper equations in the form of Pauli matrix coefficients
\begin{equation}
\begin{split}
\left(\frac{5}{3}\sigma_{0}-\sigma_{z}\right) \otimes & \left(3\sigma_{0}-\sigma_{z}\right) \\
\otimes & \left(\sigma_{0}+2\sigma_{x}-2i\sigma_{y}+\sigma_{z}\right)^{\otimes 2}
\end{split}
\end{equation}
after renormalized 
\begin{equation}
\begin{split}
\left(\frac{5}{8}\sigma_{0}-\frac{3}{8}\sigma_{z}\right) \otimes & \left(\frac{3}{4}\sigma_{0}-\frac{1}{4}\sigma_{z}\right) \\
\otimes & \left(\frac{1}{\sqrt{10}}\sigma_{0}+\frac{2}{\sqrt{10}}\sigma_{x}-\frac{2}{\sqrt{10}}i\sigma_{y}+\frac{1}{\sqrt{10}}\sigma_{z}\right)^{\otimes 2} \label{564}
\end{split}
\end{equation}
The following is a quantum gate circuit implementation of Eq.(~\ref{564}). First, the first qubit has two operations and therefore requires the introduction of an auxiliary bit. Compareded with Eq. (~\ref{546}). Make the phase gate $\sigma_ {0}=U_{0}$, $-\sigma_ {z}=U_{1}$, then for the operator $R_{y}(\theta_{1})$ can be determined by the Eq.(~\ref{546}) its angle $\theta_{1}$,
\begin{equation}
\cos^{2}\left(\frac{\theta_{1}}{2}\right)=\frac{5}{8}
\end{equation}
get $\theta_{1}=1.32$

Then there are the two operations of the second qubit that require the introduction of an auxiliary qubit. Compared with Eq.(~\ref{546}). Make the phase gate $\sigma_{0}=U_{0}$, $-\sigma_{z}=U_{1}$, then for the operator $R_{y}(\theta_{2})$ can be determined by  Eq.(~\ref{546}) its angle $\ theta _ {2} $,
\begin{equation}
\cos^{2}\left(\frac{\theta_{2}}{2}\right)=\frac{3}{4}
\end{equation}
get $\theta_{2}=\frac{\pi}{3}$

Then there are the three operations of the third qubit that require the introduction of two auxiliary qubits to implement the quantum gate $c_{0}U_{0}+c_{1}U_{1}+c_{2}U_{2}+c_{3}U_{3}$. Make the phase $\sigma_{0}=U_{0}$, $\sigma_{x}=U_{1}$, $-i\sigma_{y}=U_{2}$, $\sigma_{z}=U_{3}$

Constructing gate $R_{n}(\theta)$ according to the Eq.(~\ref{551}) 
\begin{equation}
R_{n}(\theta)|00\rangle \rightarrow \left(\frac{\sqrt{2}}{2}|0\rangle+\frac{\sqrt{2}}{2}|1\rangle\right)|0\rangle\
\end{equation}
get $R_{n}(\theta)=R_{y}(\frac{\pi}{2})$

According to the Eq.(~\ref{553}) two controlled operators $U_{1}$ and $U_{2}$ can be obtained to:
\begin{equation}
\begin{split}
U_{1}|0\rangle &\rightarrow \frac{\sqrt{5}}{5}|0\rangle+\frac{2\sqrt{5}}{5}|1\rangle\\
U_{2}|0\rangle &\rightarrow \frac{2\sqrt{5}}{5}|0\rangle+\frac{\sqrt{5}}{5}|1\rangle
\end{split}
\end{equation}
get $U_{1}=R_{y}(\theta_{3})$, $U_{2}=R_{y}(\theta_{4})$, where$\theta_{3}=2.21$, $\theta_{4}=0.93$
For the fourth working qubit and the third working qubit are the same. The quantum circuit of the operator $\Delta_{c}$ can be given as follows:
\begin{figure}[h]
\centering
\includegraphics[width=0.9\textwidth]{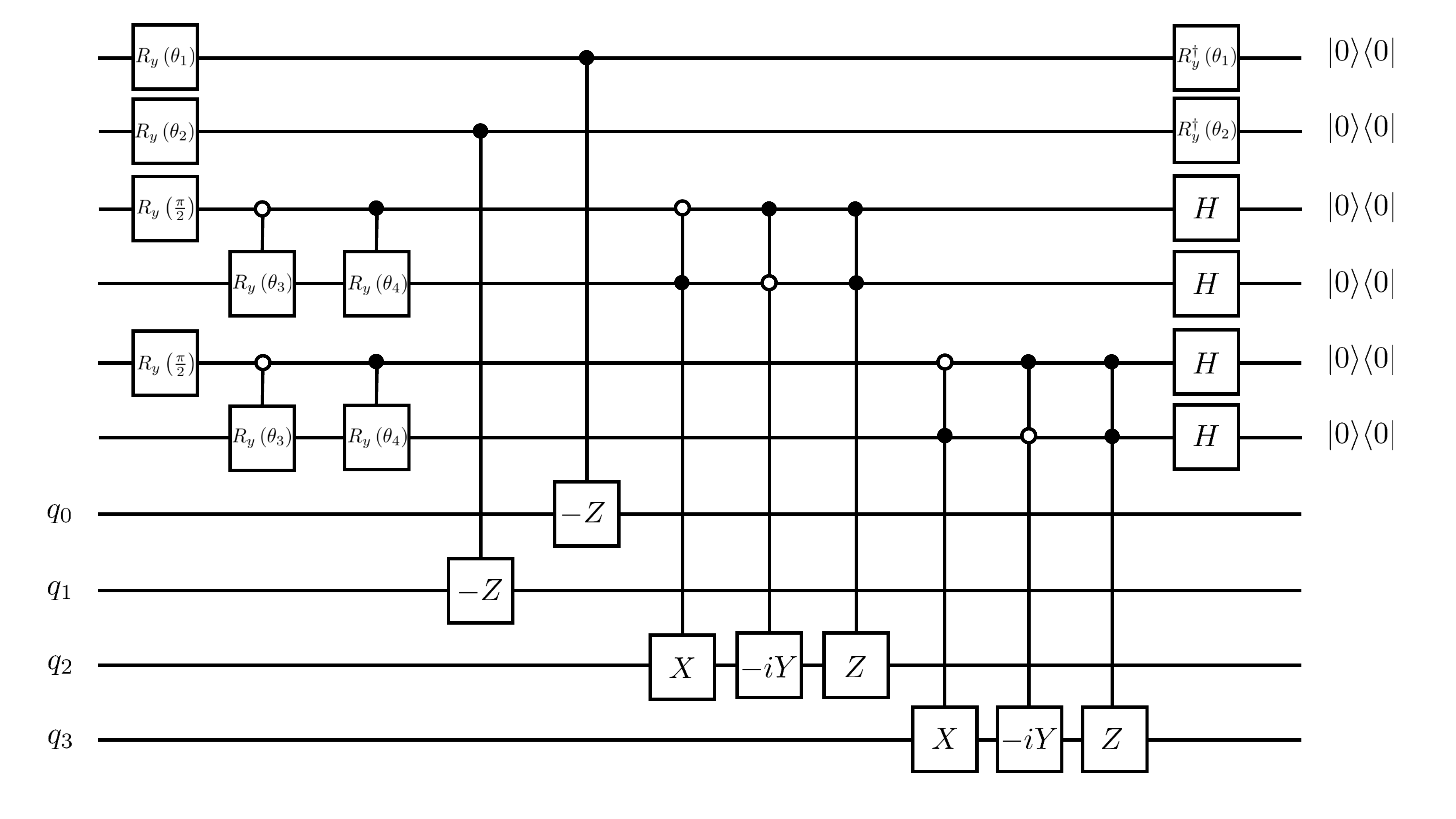}
\caption{quantum circuit of the operator $\Delta_{c}$, where $q_{0}, q_{1}, q_{2}, q_{3}$ are four working qubits, $\theta_{1}=1.32$, $\theta_{2}=\frac{\pi}{3}$ ,$\theta_{3}=2.21$, $\theta_{4}=0.93$ }
\end{figure}
The third part of the algebraic structure of Yang-Mills field theory is disassembled below.
\begin{equation}
\eta_{c}=\sum_{R} \operatorname{dim}(R) e^{-\beta C_{2}(R)}|R\rangle
\end{equation}
Write the vacuum state of the emission state:
\begin{equation}
\eta_{c}=|11\rangle\langle00|+3e^{-\frac{16}{3}i}|10\rangle \langle00|+3e^{-\frac{16}{3}i}|01\rangle \langle00|
\end{equation}
First, the superposed quantum states are disassembled into direct product forms
\begin{equation}
|1\rangle\langle0| \otimes |1\rangle\langle0| +3e^{-\frac{16}{3}i}\big[|1\rangle\langle 0|\otimes |0\rangle\langle0|+|0\rangle\langle0|\otimes |1\rangle\langle 0|\big]
\end{equation}
Expand the upper with Pauli operator bases
\begin{equation}
\begin{split}
\left(\sigma_{x}-i\sigma_{y}\right) \otimes \left(\sigma_{x}-i\sigma_{y}\right) +&3e^{-\frac{16}{3}i}\left(\sigma_{x}-i\sigma_{y}\right)\otimes \left(\sigma_{0}+\sigma_{z}\right)\\
+&3e^{-\frac{16}{3}i}\left(\sigma_{0}+\sigma_{z}\right)\otimes \left(\sigma_{x}-i\sigma_{y}\right)
\end{split}
\end{equation}
The above formula is arranged in the form of Pauli matrix coefficient
\begin{equation}
\begin{split}
\big[3e^{-\frac{16}{3}i}\sigma_{0}+\left(1+3e^{-\frac{16}{3}i}\right)\sigma_{x}&-\left(1+3e^{-\frac{16}{3}i}\right)i\sigma_{y}+3e^{-\frac{16}{3}i}\sigma_{z} \big] \\
\otimes& \left(\sigma_{0}+2\sigma_{x}-2i\sigma_{y}+\sigma_{z}\right)
\end{split}
\end{equation}
renormalized ang get
\begin{equation}
\begin{split}
\big[0.447e^{-\frac{16}{3}i}\sigma_{0}+&0.548e^{-0.73i}\sigma_{x}+0.548e^{-0.84i}\sigma_{y}+0.447e^{-\frac{16}{3}i}\sigma_{z} \big]\\
\otimes& \left(\frac{1}{\sqrt{10}}\sigma_{0}+\frac{2}{\sqrt{10}}\sigma_{x}-\frac{2}{\sqrt{10}}i\sigma_{y}+\frac{1}{\sqrt{10}}\sigma_{z}\right)
\end{split}
\end{equation}

Let's see the first qubit has three operations so we need to introduce two auxiliary qubits to implement the quantum gate $c_{0}U_{0}+c_{1}U_{1}+c_{2}U_{2}+c_{3}U_{3}$ where $e^{-\frac{16}{3}i}\sigma_{0}=U_{0}$, $e^{-0.73i}\sigma_{x}=U_{1}$, $e^{-0.84i}\sigma_{y}=U_{2}$, $e^{-\frac{16}{3}i}\sigma_{z}=U_{3}$ 
The effect of gate $R _ {n} (\ theta) $ according to the Eq.(~\ref{551}) is:
\begin{equation}
R_{n}(\theta)|00\rangle \rightarrow \left(\frac{\sqrt{2}}{2}|0\rangle+\frac{\sqrt{2}}{2}|1\rangle\right)|0\rangle\
\end{equation}
get $R_{n}(\theta)$ is $R_{y}(\frac{\pi}{2})$

According to the Eq.(~\ref{553}) two controlled operators $U_{1}$  and $U_{2}$ can be obtained to: 
\begin{equation}
\begin{split}
U_{1}|0\rangle &\rightarrow 0.632|0\rangle+0.775|1\rangle\\
U_{2}|0\rangle &\rightarrow 0.775|0\rangle+0.642|1\rangle
\end{split}
\end{equation}
get $U_{1}=R_{y}(\theta_{1})$, $U_{2}=R_{y}(\theta_{2})$, where $\theta_{1}=1.77$, $\theta_{2}=1.37$ 
Let's see the second qubit has three operations so we need to introduce two auxiliary qubits to implement the quantum gate$c_{0}U_{0}+c_{1}U_{1}+c_{2}U_{2}+c_{3}U_{3}$ where $\sigma_{0}=U_{0}$, $\sigma_{x}=U_{1}$, $-i\sigma_{y}=U_{2}$, $\sigma_{z}=U_{3}$ 

The effect of constructing gate $R_{n}(\theta)$ according to the Eq.(~\ref{551}) is:
\begin{equation}
R_{n}(\theta)|00\rangle \rightarrow \left(\frac{\sqrt{2}}{2}|0\rangle+\frac{\sqrt{2}}{2}|1\rangle\right)|0\rangle\
\end{equation}
get $R_{n}(\theta)$ is $R_{y}(\frac{\pi}{2})$

According to the Eq.(~\ref{553}) two controlled operators $U_{1}$ and $U_{2}$ can be obtained to:
\begin{equation}
\begin{split}
U_{1}|0\rangle &\rightarrow \frac{\sqrt{5}}{5}|0\rangle+\frac{2\sqrt{5}}{5}|1\rangle\\
U_{2}|0\rangle &\rightarrow \frac{2\sqrt{5}}{5}|0\rangle+\frac{\sqrt{5}}{5}|1\rangle
\end{split}
\end{equation}
get $U_{1}=R_{y}(\theta_{3})$, $U_{2}=R_{y}(\theta_{4})$, where$\theta_{3}=2.21$, $\theta_{4}=0.93$ The quantum circuit of the operator $\eta_{c}$ can be given as follows:
\begin{figure}[h]
\centering
\includegraphics[width=0.9\textwidth]{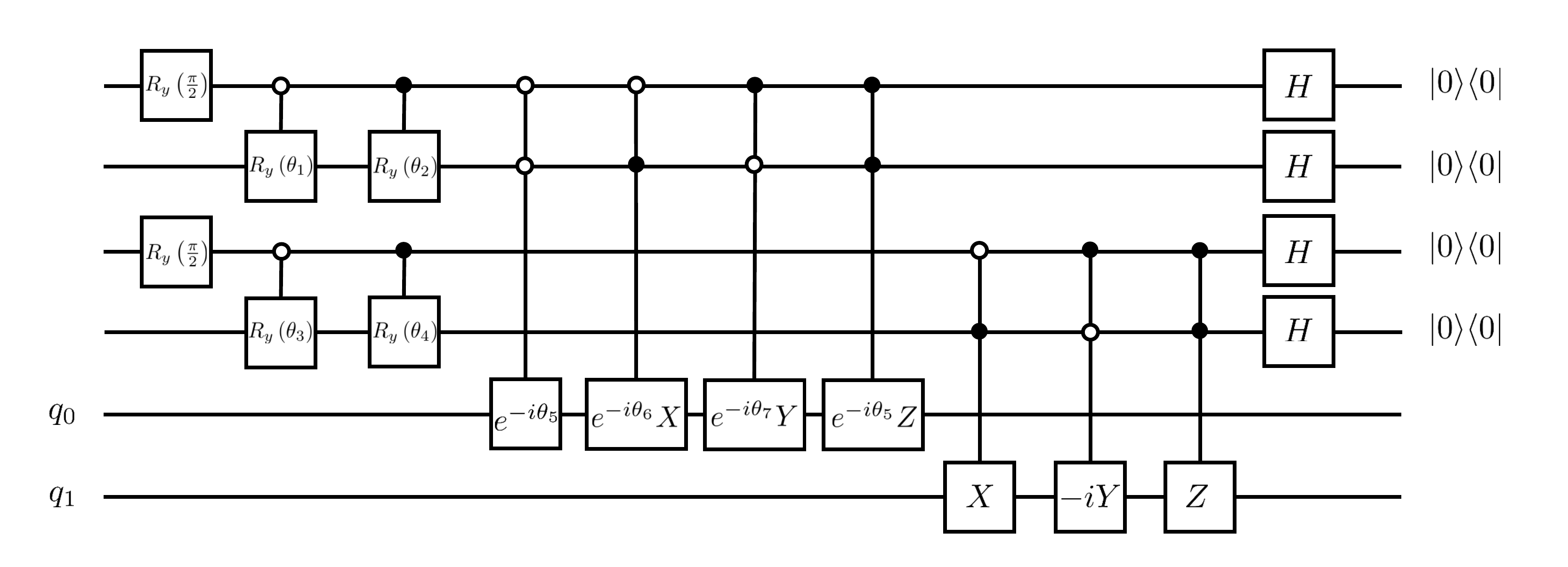}
\caption{Quantum circuit of the operator $\eta_{c}$,where $q_{0}, q_{1}$ are two working qubit $\theta_{1}=1.77$, $\theta_{2}=1.37$ , $\theta_{3}=2.21$, $\theta_{4}=0.93$, $\theta_{5}=\frac{16}{3}$, $\theta_{6}=0.73$, $\theta_{7}=0.84$ }
\end{figure} 

Below is the dismantling of the fourth part of Yang-Mills field theory.
\begin{equation}
\epsilon_{c}=\sum_{R} \mathrm{dim}(R) \langle R|
\end{equation}
Write the vacuum state of the incident state:
\begin{equation}
\epsilon_{c}=|00\rangle\langle11|+3|00\rangle \langle 10|+3|00\rangle \langle01|
\end{equation}
First, the superposed quantum states are disassembled into direct product forms
\begin{equation}
|0\rangle\langle 1| \otimes |0\rangle\langle 1| +3\big[|0\rangle\langle 1|\otimes |0\rangle\langle0|+|0\rangle\langle0|\otimes |0\rangle\langle 1|\big]
\end{equation}
Expand the upper using Pauli bases
\begin{equation}
\begin{split}
\left(\sigma_{x}+i\sigma_{y}\right) \otimes \left(\sigma_{x}+i\sigma_{y}\right) +&3\left(\sigma_{x}+i\sigma_{y}\right)\otimes \left(\sigma_{0}+\sigma_{z}\right)\\
+&3\left(\sigma_{0}+\sigma_{z}\right)\otimes \left(\sigma_{x}+i\sigma_{y}\right)
\end{split}
\end{equation}
The above formula is arranged in the form of Pauli matrix coefficient
\begin{equation}
\big[3\sigma_{0}+4\sigma_{x}+4i\sigma_{y}+3\sigma_{z} \big] 
\otimes \left(\sigma_{0}+2\sigma_{x}+2i\sigma_{y}+\sigma_{z}\right)
\end{equation}
After renormalizing
\begin{equation}
\begin{split}
\Bigg(\frac{3 \sqrt{2}}{10}\sigma_{0}+&\frac{2\sqrt{2}}{5}\sigma_{x}+\frac{2\sqrt{2}}{5}\sigma_{y}+\frac{3\sqrt{2}}{10}\sigma_{z} \Bigg)\\
\otimes& \left(\frac{1}{\sqrt{10}}\sigma_{0}+\frac{2}{\sqrt{10}}\sigma_{x}+\frac{2}{\sqrt{10}}i\sigma_{y}+\frac{1}{\sqrt{10}}\sigma_{z}\right)
\end{split}
\end{equation}
Let's see the first qubit has three operations so we need to introduce two auxiliary qubits to implement the quantum gate $c_{0}U_{0}+c_{1}U_{1}+c_{2}U_{2}+c_{3}U_{3}$ ,where $\sigma_{0}=U_{0}$, $\sigma_{x}=U_{1}$, $\sigma_{y}=U_{2}$, $\sigma_{z}=U_{3}$ 
The effect of gate $R _ {n} (\ theta) $ according to the Eq.(~\ref{551}) is:
\begin{equation}
R_{n}(\theta)|00\rangle \rightarrow \left(\frac{\sqrt{2}}{2}|0\rangle+\frac{\sqrt{2}}{2}|1\rangle\right)|0\rangle\
\end{equation}
get $R_{n}(\theta)$ is $R_{y}(\frac{\pi}{2})$

According to the Eq.(~\ref{553}) two controlled operators $U_{1}$ and $U_{2}$ can be obtained to:
\begin{equation}
\begin{split}
U_{1}|0\rangle &\rightarrow \frac{3}{5}|0\rangle+\frac{4}{5}|1\rangle\\
U_{2}|0\rangle &\rightarrow \frac{4}{5}|0\rangle+\frac{3}{5}|1\rangle
\end{split}
\end{equation}
get $U_{1}=R_{y}(\theta_{1})$, $U_{2}=R_{y}(\theta_{2})$, where $\theta_{1}=1.85$, $\theta_{2}=1.29$ 
Let's see the second qubit has three operations so we need to introduce two auxiliary qubits to implement the quantum gate$c_{0}U_{0}+c_{1}U_{1}+c_{2}U_{2}+c_{3}U_{3}$, where $\sigma_{0}=U_{0}$, $\sigma_{x}=U_{1}$, $i\sigma_{y}=U_{2}$, $\sigma_{z}=U_{3}$ 
The effect of gate $R _ {n} (\ theta) $ according to the Eq.(~\ref{551}) is:
\begin{equation}
R_{n}(\theta)|00\rangle \rightarrow \left(\frac{\sqrt{2}}{2}|0\rangle+\frac{\sqrt{2}}{2}|1\rangle\right)|0\rangle\
\end{equation}
get $R_{n}(\theta)$is $R_{y}(\frac{\pi}{2})$

According to the Eq.(~\ref{553}) two controlled operators $U_{1}$ and $U_{2}$ can be obtained to:
\begin{equation}
\begin{split}
U_{1}|0\rangle &\rightarrow \frac{\sqrt{5}}{5}|0\rangle+\frac{2\sqrt{5}}{5}|1\rangle\\
U_{2}|0\rangle &\rightarrow \frac{2\sqrt{5}}{5}|0\rangle+\frac{\sqrt{5}}{5}|1\rangle
\end{split}
\end{equation}
get $U_{1}=R_{y}(\theta_{3})$, $U_{2}=R_{y}(\theta_{4})$, where$\theta_{3}=2.21$, $\theta_{4}=0.93$ The quantum circuit of the operator $\eta_{c}$ can be expressed as follows:
\begin{figure}[h]
\centering
\includegraphics[width=0.9\textwidth]{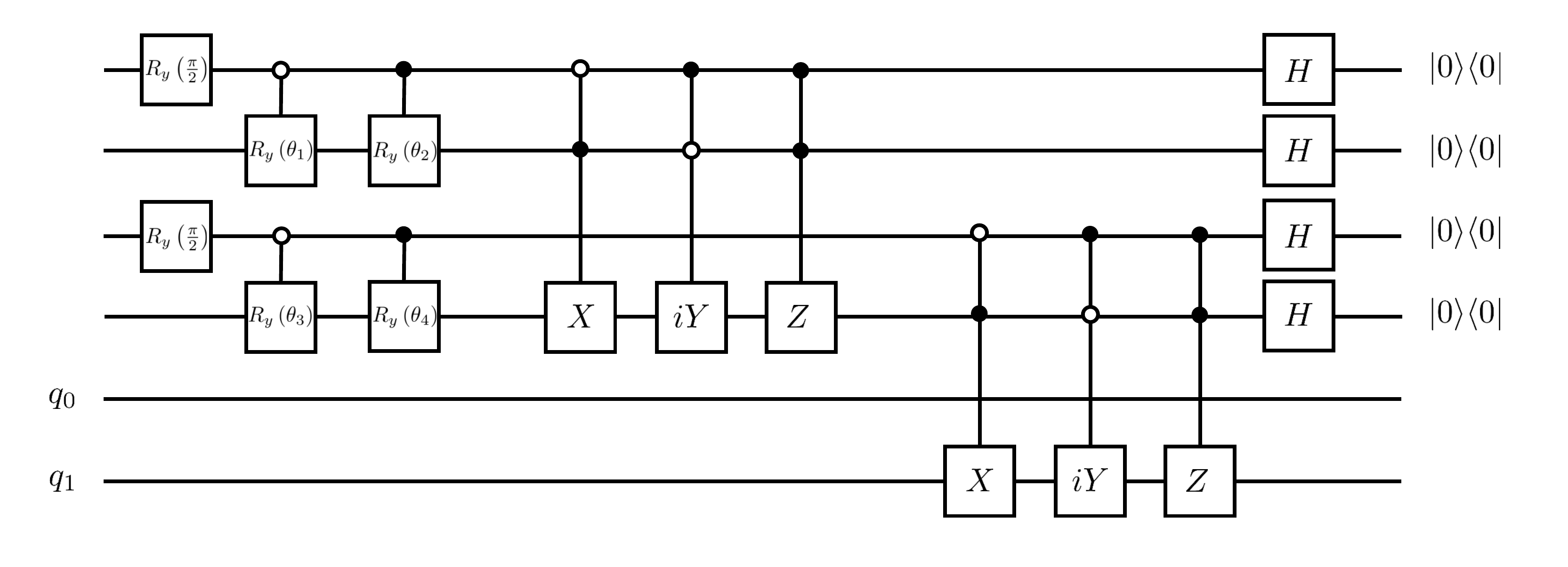}
\caption{Quantum circuit of the operator $\eta_{c}$, where$q_{0}, q_{1}$ are two working qubit  $\theta_{1}=1.85$, $\theta_{2}=1.29$ , $\theta_{3}=2.21$, $\theta_{4}=0.93$ }
\end{figure}

\begin{figure}[h]
\centering
\includegraphics[width=1.1\textwidth]{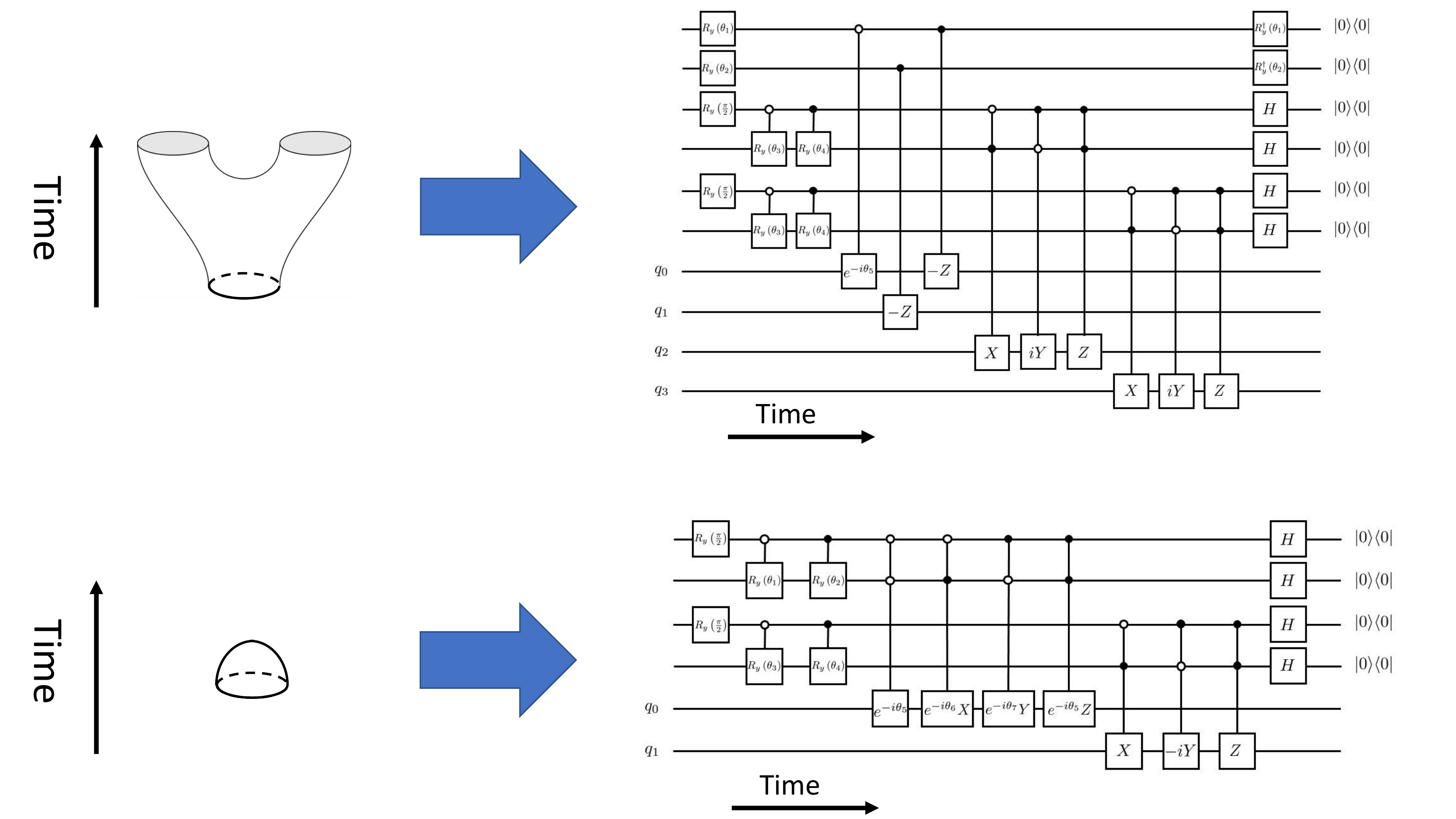}
\quad
\\
\quad
\includegraphics[width=1.1\textwidth]{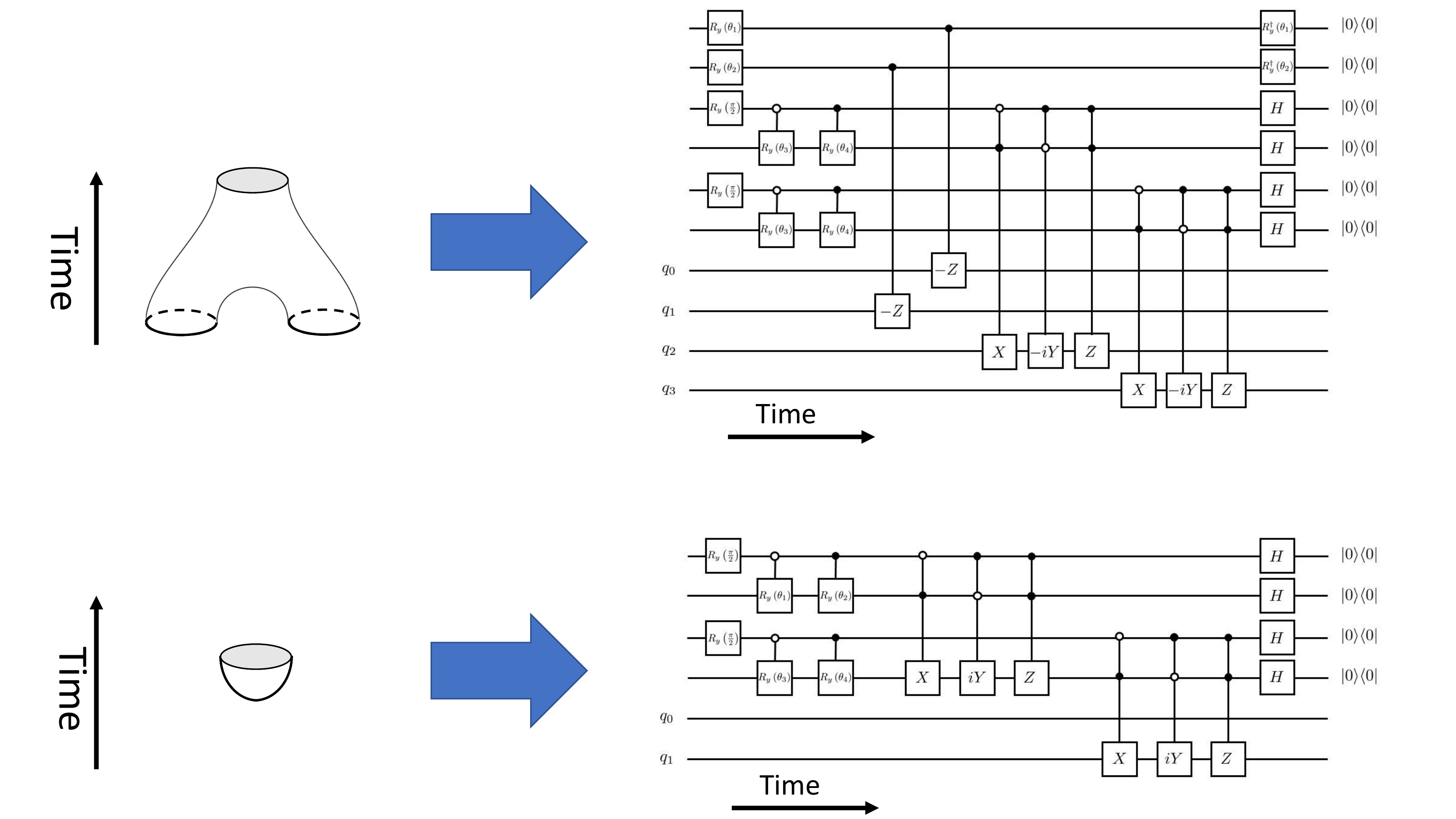}
\caption{Catigorical quantum simulation of $SU(3)$ Yang-Mills theory}
\label{ymtf}
\end{figure}

\paragraph{Discussion} At this point, the author completes a categorical simulation of the $SU(3)$ Yang-Mills theory. The author collates the final results into Figure~\ref{ymtf}. On the left side of the diagram is the state emission of the $\mathbf{Bord}_{2}$ and on the right is the state emission of the $\mathbf{QC}$ of the quantum circuit category. The blue arrow represents functor $F$. Figure~\ref{ymtf} precisely the process described by Eq.(~\ref{fclzmn}) ~. In the mathematical sense, categorical quantum simulation is a binary quantum field theory. Because it's a process of mapping geometric categories to quantum circuit categories. In combination with the introduction of TQFT , Figure~\ref{ymtf}, the $SU (3) $ Yang-Mills theory is determined entirely by the bordism category (equivalent to the commutative of Frobenius algebra category). And Figure~\ref{ymtf} completely maps the bordism category to the quantum circuit category  via the function $F$. Objects in the bordism category (non-intersecting circles) are mapped to objects in the quantum circuit category (qubits $q_{0}-q_{3}$) by the  encoding of Fifure~\ref{bm}, and the morphism of the bordism category are mapped to the operation of the quantum circuit (that is what Figure~\ref{ymtf} shown). In topological quantum field theory, all propagons can be disassembled into the composition of the morphism in bordism category. So,  connect the quantum circuit in the Figure~\ref{ymtf}~\ref{ymtf} will simulate the Yang-Mills field.

Using quantum simulations of the Feynman Paradigm to accomplish the same task would require a lot of unnecessary bit resources. Because traditional quantum simulations require the dispersion of space, encoding the states of each point in space to the quantum states of quantum computers. In contrast, categorical quantum simulations encode Hilbert space directly, considering its algebraic structural space rather than geometric space. Therefore, categorical quantum simulations saves qubit resources.

Figure~\ref{ymtf} also reveals the profound physics behind the $SU(3)$ Yang-Mills theory. It can be seen that the quantum circuit in Figure ~\ref{ymtf} use the duality pattern of quantum computing. That is, by introducing auxiliary bits. This reveals that the field in the $SU(3)$ Yang-Mills theory is subject to a gravitational norm anomaly (anomaly refers to a $n$ dimensional quantum field that cannot be achieved on a $n$ lattice model). This shows that the $SU(3)$ Yang-Mills theory is extremely long range entanglement. To entangle low dimension space through higher dimensions.

\bibliographystyle{apsrev4-1}
\bibliography{refs}

\end{document}